\input ./style/arxiv-general.cfg
\documentclass[aos,MSNbibl,nameyear,dvips]{arximspdf}
\makeatletter
   \@ifpackageloaded{graphicx}{}{\usepackage{graphicx}}
\makeatother
\usepackage{mathbh}
\usepackage{algorithm,algorithmicx}

\doi{10.1214/15-AOS1373}
\volume{44}
\issue{2}
\pubyear{2016}
\firstpage{489}
\lastpage{514}
\docsubty{FLA}

\makeatletter
\def\sfrac#1#2{#1/#2}
\def\vfrac#1#2{(#1)/#2}

\renewcommand{\mid}{|}
\newcommand{\rrvert}{\vert}
\newcommand{\rrVert}{\Vert}
\newcommand{\llvert}{\vert}
\newcommand{\llVert}{\Vert}
\newtheorem{teo}{Theorem}
\newtheorem{lem}[teo]{Lemma}
\newtheorem{cor}[teo]{Corollary}
\newcommand{\eqref}[1]{(\ref{#1})}
\let\hat\widehat
\let\tilde\widetilde
\newcommand{\R}{\mathbb{R}}
\renewcommand\P{\mathbb{P}}
\newcommand{\MISE}{{\sf MISE}}
\newcommand{\dest}{{\sf dest}}
\newcommand{\length}{{\sf length}}
\newcommand{\Haus}{{\sf Haus}}
\newcommand{\Vol}{{\sf Vol}}
\newcommand{\Quantile}{{\sf Quantile}}
\newcommand{\K}{D}
\newcommand{\Cov}{\operatorname{Cov}}
\makeatother

\begin{document}
\begin{frontmatter}

\title{Nonparametric modal regression}
\runtitle{Nonparametric modal regression}

\begin{aug}
\author[A]{\fnms{Yen-Chi}~\snm{Chen}\corref{}\thanksref{T1}\ead[label=e1]{yenchic@andrew.cmu.edu}},
\author[A]{\fnms{Christopher R.}~\snm{Genovese}\thanksref{T2}\ead[label=e2]{genovese@cmu.edu}},
\author[A]{\fnms{Ryan~J.}~\snm{Tibshirani}\thanksref{T3}\ead[label=e3]{ryantibs@cmu.edu}}
\and
\author[A]{\fnms{Larry}~\snm{Wasserman}\thanksref{T4}\ead[label=e4]{larry@cmu.edu}}
\runauthor{Chen, Genovese, Tibshirani and Wasserman}
\affiliation{Carnegie Mellon University}
\address[A]{Department of Statistics\\
Carnegie Mellon University\\
5000 Forbes Ave.\\
Pittsburgh, Pennsylvania 15213\\
USA\\
\printead{e1}\\
\phantom{E-mail: }\printead*{e2}\\
\phantom{E-mail: }\printead*{e3}\\
\phantom{E-mail: }\printead*{e4}}
\end{aug}
\thankstext{T1}{Supported by DOE Grant DE-FOA-0000918.}
\thankstext{T2}{Supported in part by DOE Grant DE-FOA-0000918 and NSF Grant DMS-1208354.}
\thankstext{T3}{Supported by NSF Grant DMS-13-09174.}
\thankstext{T4}{Supported by NSF Grant DMS-12-08354.}

%
\received{\smonth{12} \syear{2014}}
%
\revised{\smonth{8} \syear{2015}}

%
\begin{abstract}
Modal regression estimates the local modes of the distribution of $Y$
given $X=x$, instead of the mean, as in the usual regression sense,
and can hence reveal important structure missed by usual
regression methods. We study a simple nonparametric method for modal
regression, based on a kernel density estimate (KDE) of the joint
distribution of $Y$ and $X$. We derive asymptotic error bounds for
this method, and propose techniques for constructing confidence
sets and prediction sets. The latter is used to select the
smoothing bandwidth of the underlying KDE. The idea behind modal
regression is connected to many others, such as mixture regression
and density ridge estimation, and we discuss these ties as well.
\end{abstract}

%
\end{frontmatter}

\section{Introduction}
\label{secintro}

Modal regression [\citet{sager1982maximum,Lee1989,Yao2012,yao2014new}]
is an alternate approach to the usual regression methods for
exploring the relationship between a response variable $Y$ and
a predictor variable $X$.
Unlike conventional regression, which is based on the conditional mean
of $Y$ given $X=x$, modal regression estimates conditional modes of
$Y$ given $X=x$.

\begin{figure}

\includegraphics{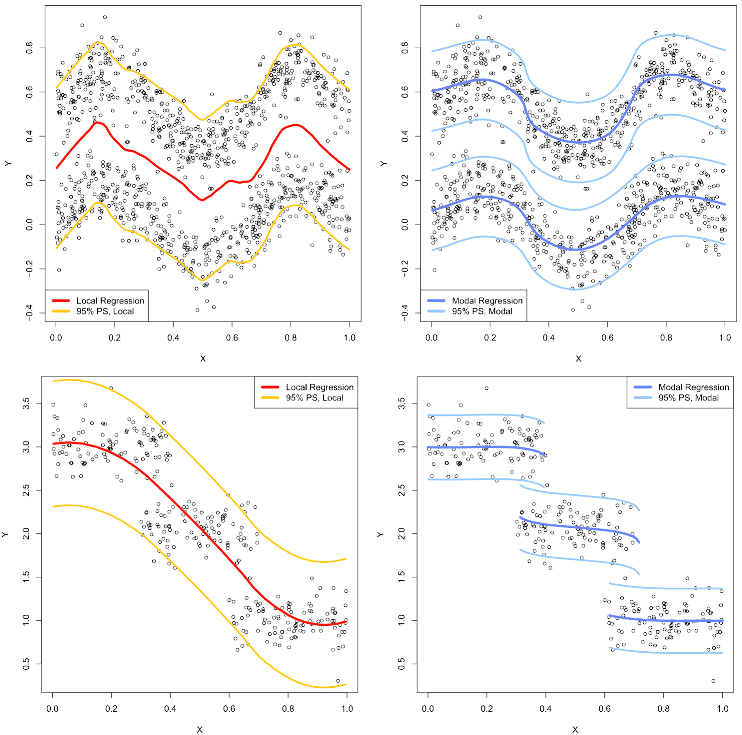}

\caption{Examples of modal regression versus a common nonparametric
regression estimator, local linear regression. In the top row, we
show local regression estimate and its associated 95\% prediction
bands alongside the modal regression and its 95\% prediction bands.
The bottom row does the same for a different data example. The
local regression method fails to capture the structure, and produces
prediction bands that are too wide.}

\label{Figexample1}
\end{figure}

Why would we ever use modal regression in favor
a conventional regression method? The answer, at a high-level, is that
conditional modes can reveal structure that is missed by the
conditional mean. Figure~\ref{Figexample1} gives a definitive
illustration of this point: we can see that, for the data examples in
question, the conditional mean both fails to
capture the major trends present in the response, and produces
unnecessarily wide prediction bands. Modal regression is an
improvement in both of these regards (better trend estimation and
narrower prediction bands). In this paper, we rigorously study and
develop its properties.

%

Modal regression has been used in transportation
[\citet{Einbeck2006}], astronomy [\citet{Rojas2005}], meteorology
[\citet{Hyndman1996}] and economics
[\citet{Huang2012jasa,Huang2013jasa}].
Formally, the conditional (or local) mode set at $x$ is
defined as
%
\begin{equation}
\label{eqM0} M(x) = \biggl\{y: \frac{\partial}{\partial y} p(y\mid x)=0,
\frac{\partial^2}{\partial y^2} p(y\mid x)<0 \biggr\},
\end{equation}
where $p(y\mid x) = p(x,y)/p(x)$ is the conditional density of $Y$ given
$X=x$. As a simplification, the set $M(x)$ can be
expressed in terms of the joint density alone:
%
\begin{equation}
\label{eqM1} M(x) = \biggl\{y: \frac{\partial}{\partial y} p(x,y)=0,
\frac{\partial^2}{\partial y^2}
p(x,y)<0 \biggr\}.
\end{equation}
%
At each $x$, the local mode set $M(x)$ may
consist of several points, and so $M(x)$ is in general a
multivalued function.
Under appropriate conditions, as we will show, these modes change
smoothly as $x$ changes. Thus, local modes behave like a
collection of surfaces which we call \emph{modal manifolds}.

We focus on a nonparametric estimate of the conditional mode set,
derived from a kernel density estimator (KDE):
%
\begin{equation}
\label{eqMhat1} \hat{M}_n(x) = \biggl\{y: \frac{\partial}{\partial y}
\hat{p}_n(x,y)=0, \frac{\partial^2}{\partial y^2} \hat{p}_n(x,y)<0
\biggr\},
\end{equation}
where
$\hat{p}_n(x,y)$ is the joint KDE of $X,Y$.
\citet{scott1992multivariate} proposed this plug-in estimator
for modal regression, and \citet{Einbeck2006} proposed a fast
algorithm. We extend the work of these authors by
giving a thorough treatment and analysis of nonparametric modal
regression. In particular, our contributions are as follows.
\begin{longlist}[2.]
\item[1.] We study the geometric properties of modal regression.
\item[2.] We prove consistency of the nonparametric modal regression
estimator, and furthermore derive explicit convergence rates, with
respect to various error metrics.
\item[3.] We propose a method for constructing confidence sets, using
the bootstrap, and prove that it has proper asymptotic coverage.
\item[4.] We propose a method for constructing prediction sets, based on
plug-in methods, and prove that the population prediction sets
from this method can be smaller than those based on the regression
function.
\item[5.] We propose a rule for selecting the smoothing bandwidth of the
KDE 
based on
minimizing the size of prediction sets.
\item[6.] We draw enlightening comparisons to mixture regression (which
suggests a clustering method using modal regression) and to
density ridge estimation.
\end{longlist}

We begin by reviewing basic properties of modal regression and
recalling previous work, in Section~\ref{secreview}. Sections~\ref
{secgeometry} through \ref{secdensity} then follow roughly
the topics described in items 1--6 above. In Section~\ref
{secdiscussion}, we end with some discussion.
Simple R code for the modal regression and some simulation data sets
used in this paper can be found at
\surl{http://www.stat.cmu.edu/\textasciitilde yenchic/\\ModalRegression.zip}.

\section{Review of modal regression}\label{secreview}

Consider a response variable $Y \in\mathbb{K}\subseteq\R$ and
covariate or predictor variable $X \in\K\subseteq\R^d$, where $\K$
is a compact set. A~classic take on modal regression
[\citet{sager1982maximum,Lee1989,yao2014new}] is
to assume a linear model
\[
{\sf Mode}(Y\mid X=x) = \beta_0 + \beta^T x,
\]
where $\beta_0 \in\R$, $\beta\in\R^d$ are unknown coefficients, and
${\sf Mode}(Y\mid X=x)$ denotes the (global) mode of $Y$ given $X=x$.
Nonparametric modal regression is more flexible,
because it allows $M(x)$ to be multivalued, and also it models the
components of $M(x)$ as smooth functions of $x$ (not necessarily
linear). As another nonlinear generalization of the above model,
\citet{Yao2012} propose an interesting local polynomial smoothing
method for mode estimation; however, they focus on the global mode
${\sf Mode}(Y\mid X=x)$ rather than $M(x)$, the collection of all
conditional modes.

The estimated local mode set {$\hat{M}_n(x)$} in
\eqref{eqMhat1} from \citet{scott1992multivariate} relies on an
estimated joint density function
{$\hat{p}_n(x,y)$}, most commonly computed using a KDE. Let
$(X_1,Y_1),\ldots,(X_n,Y_n)$ be the observed data samples. Then the
KDE of the joint density $p(x,y)$ is
%
\begin{equation}
\label{eqKDE} \hat{p}_n(x,y) = \frac{1}{nh^{d+1}}\sum
_{i=1}^n K \biggl(\frac{\llVert x-X_i\rrVert}{h} \biggr) K
\biggl( \frac{y-Y_i}{h} \biggr).
\end{equation}
Here, $K$ is a symmetric, smooth kernel function, such as the Gaussian
kernel [i.e., $K(u)=e^{-u^2/2}/\sqrt{2\pi}$], and $h>0$ is
the smoothing bandwidth. For simplicity, we have used the same
kernel function $K$ and bandwidth $h$ for the covariates and the
response, but this is not necessary. For brevity, we will write the
estimated modal set as
%
\begin{equation}
\label{eqMhat2} \hat{M}_n(x) = \bigl\{y: \hat{p}_{y,n}(x,y) =
0, \hat{p}_{yy,n}(x,y) <0 \bigr\},
\end{equation}
where the subscript notation denotes partial derivatives, as in $f_y
= \partial f(x,y)/\partial y$ and $f_{yy} = \partial^2 f(x,y)/\partial
y^2$.

In general, calculating {$\hat{M}_n(x)$} can be
challenging, but for special kernels, \citet{Einbeck2006}
proposed a simple and efficient algorithm for computing local mode
estimates, based on the mean-shift algorithm
[\citet{cheng1995mean,comaniciu2002mean}].
A related approach can be found in \citet{yao2013note}, where the
authors consider a mode hunting algorithm based on the EM algorithm.
For a discussion of how the mean-shift and EM algorithms
relate, see \citet{carreira2007gaussian}.
In general, mean-shift algorithms can be derived for any KDEs with
radially symmetric kernels [\citet{comaniciu2002mean}], but for
simplicity we assume here that $K$ is Gaussian. The partial
mean-shift algorithm of \citet{Einbeck2006}, to estimate conditional
modes, is described in Algorithm \ref{algPMS}.

\begin{algorithm}
\caption{Partial mean-shift}
\begin{algorithmic}
\State\textbf{Input:} Data samples
$\mathcal{D}=\{(X_1,Y_1),\ldots,(X_n,Y_n)\}$,
bandwidth $h$. (The kernel $K$ is chosen to be Gaussian.)
\begin{enumerate}
\item Initialize mesh points $\mathcal{M}\subseteq\R^{d+1}$ (a
common choice is $\mathcal{M}=\mathcal{D}$).
\item For each $(x,y)\in\mathcal{M}$, fix $x$, and update $y$ using
the following iterations until convergence:
%
\begin{equation}
y \longleftarrow\frac{
\sum_{i=1}^n Y_i K (\sfrac{\llVert x-X_i\rrVert
}{h} )
K ( \vfrac{y-Y_i}{h} )}{
\sum_{i=1}^n K (\sfrac{\llVert x-X_i\rrVert}{h} )
K (\vfrac{y-Y_i}{h} )}. \label{eqms}
\end{equation}
\end{enumerate}
\State\textbf{Output:} The set $\mathcal{M}^\infty$, containing the
points $(x,y^\infty)$, where $x$ is a predictor value as fixed in
$\mathcal{M}$, and $y^\infty$ is the corresponding limit of the
mean-shift iterations \eqref{eqms}.
\end{algorithmic}
\label{algPMS}
\end{algorithm}

A straightforward calculation shows that the mean-shift update
\eqref{eqms} is indeed a gradient ascent update on the function
{$f(y) = \hat{p}_n(x,y)$} (for fixed $x$), with an implicit
choice of step size. Because this function
$f$ is generically nonconcave, we are not guaranteed that gradient
ascent will actually attain a (global) maximum, but it will converge
to critical points under small enough step sizes
[\citet{arias2013estimation}].

\section{Geometric properties}\label{secgeometry}

\begin{figure}[t]

\includegraphics{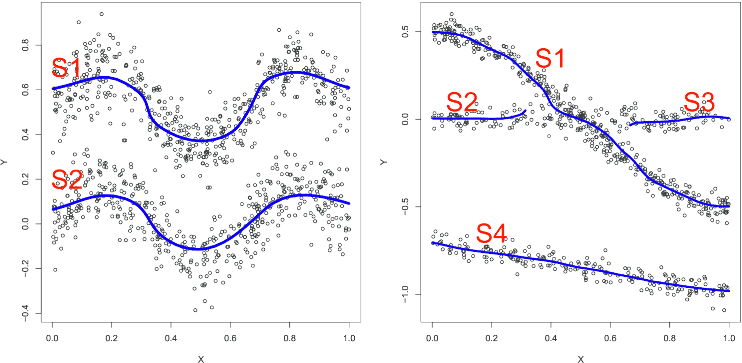}

\caption{Examples of modal manifolds.}
\label{Figthm1}
\end{figure}

In this section, we study the geometric properties of modal
regression. Recall that $M(x)$ is a set of points at each input $x$.
We define the \emph{modal manifold collection} as the union of these
sets over all inputs $x$,
%
\begin{equation}
\mathcal{S} = \bigl\{ (x,y): x\in\K, y\in M(x)\bigr\}. \label{eqS0}
\end{equation}
By the implicit function theorem, the set $\mathcal{S}$ has dimension
$d$; see Figure~\ref{Figthm1} for an illustration with $d=1$
(univariate $x$).

%

We will assume that the modal manifold collection
$\mathcal{S}$ can be factorized as
%
\begin{equation}
\label{eqS1} \mathcal{S} = S_1\cup\cdots\cup S_K,
\end{equation}
where each $S_j$ is a connected manifold
that admits a parametrization
%
\begin{equation}
S_j = \bigl\{\bigl(x,m_j(x)\bigr): x\in
A_j\bigr\}, \label{eqS2}
\end{equation}
for some function $m_j(x)$ and open set $A_j$.
For instance, in Figure~\ref{Figthm1},
each $S_j$ is a connected curve.
Note that
$A_1,\ldots,A_K$ form an open cover for the support $\K$ of $X$.
We call $S_j$ the \emph{$j$th modal manifold}, and
$m_j(x)$ the \emph{$j$th modal function}.
By convention, we let $m_j(x)=\varnothing$ if $x\notin A_j$ and,
therefore, we may write
%
\begin{equation}
\label{eqMFdef} M(x) = \bigl\{m_1(x),\ldots,m_K(x)
\bigr\}.
\end{equation}
That is, at any $x$, the values among $m_1(x),\ldots,m_K(x)$
that are nonempty give local modes.

Under weak assumptions, each $m_j(x)$ is differentiable,
and so is the modal set $M(x)$, in a sense. We discuss this next.

%
\begin{lem}[(Derivative of modal functions)]\label{lemcurves}
Assume that $p$ is twice differentiable, and
let $\mathcal{S}=\{(x,y): x \in\K, y \in M(x)\}$ be the
modal manifold collection. Assume that $\mathcal{S}$ factorizes
according to \eqref{eqS0}, \eqref{eqS1}. Then, when $x \in A_j$,
%
\begin{equation}
\label{eqmjgrad} \nabla m_j(x) = -\frac{p_{yx}(x,m_j(x))}{p_{yy}(x,m_j(x))},
\end{equation}
where
$p_{yx}(x,y) = \nabla_x \frac{\partial}{\partial y} p(x,y)$ is the
gradient over $x$ of $p_y(x,y)$.
\end{lem}

\begin{pf}
Since we assume that $x\in A_j$, we have
$p_y(x,m_j(x))=0$ by definition. Taking a gradient over $x$ yields
\[
0 = \nabla_x p_{y}\bigl(x,m_j(x)\bigr) =
p_{yx}\bigl(x,m_j(x)\bigr) + p_{yy}
\bigl(x,m_j(x)\bigr)\nabla m_j(x). %
\]
After rearrangement,
\[
\nabla m_j(x) = -\frac{p_{yx}(x,m_j(x))}{p_{yy}(x,m_j(x))}. %
\]\upqed
\end{pf}

Lemma \ref{lemcurves} links the geometry of the joint density
function
to the smoothness of the modal functions (and modal manifolds).
The formula \eqref{eqmjgrad} is well-defined as long as
$p_{yy}(x,m_j(x))$ is nonzero, which is guaranteed by the definition
of local modes. Thus, when $p$ is smooth, each modal manifold is
also smooth.

To characterize smoothness of $M(x)$ itself, we require a notion for
smoothness over sets. For this, we recall the
\emph{Hausdorff distance} between two sets $A,B$, defined as
%
\[
\Haus(A,B) = \inf\{r: A\subseteq B\oplus r, B\subseteq A\oplus r\}, %
\]
%
where $A\oplus r= \{x: d(x,A)\leq r\}$ with
$d(x,A)= \inf_{y\in A}\llVert x-y\rrVert$.

%
\begin{teo}[(Smoothness of the modal manifold collection)]\label{teosmooth}
Assume the conditions of Lemma \ref{lemcurves}. Assume furthermore
all partial derivatives of $p$ are bounded by $C$, and
there exists $\lambda_2>0$ such that $p_{yy}(x,y)<-\lambda_2$ for all
$y\in M(x)$ and $x\in\K$. Then
%
\begin{equation}
\label{eqsmooth} \lim_{\llvert\varepsilon\rrvert\rightarrow0}\frac
{\Haus(M(x),
M(x+\varepsilon))}{\llvert\varepsilon\rrvert} \leq\max
_{j=1,\ldots, K} \bigl\llVert m'_j(x)\bigr
\rrVert\leq\frac{C}{\lambda_2} < \infty.
\end{equation}
\end{teo}

The proof of this result follows directly from Lemma~\ref{lemcurves}
and the definition of Hausdorff distance, so we omit it. Since $M(x)$
is a multivalued function, classic notions of smoothness cannot be
applied, and Theorem~\ref{teosmooth} describes its smoothness
in terms of Hausdorff distance. This distance can be thought of as
a generalized $\ell_\infty$ distance for sets, and
Theorem~\ref{teosmooth} can be interpreted as a statement about
Lipschitz continuity with respect to Hausdorff distance.

%
\begin{figure}
\begin{tabular}{@{}cc@{}}

\includegraphics{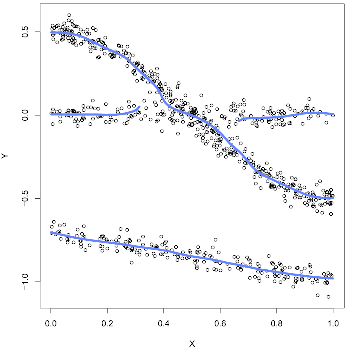}  & \includegraphics{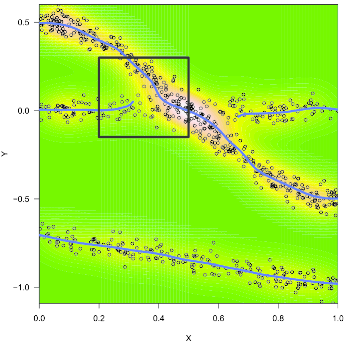}\\
\footnotesize{(a) Modal regression} & \footnotesize{(b) Joint density contour}\\[6pt]

\includegraphics{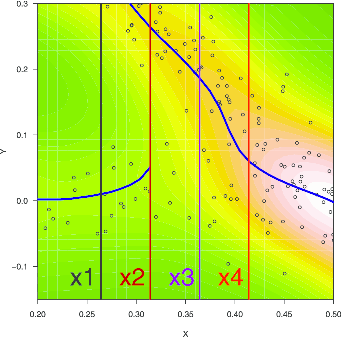}  & \includegraphics{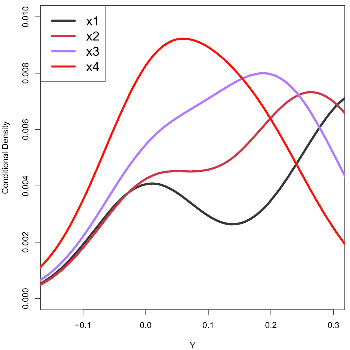}\\
\footnotesize{(c) Zoomed-in density contour} & \footnotesize{(d) Conditional density given $X=x_i$}\\
\end{tabular}
\caption{A look at bifurcation. As $X$ moves $x_1$ to $x_4$, we can
see that a local mode disappears after $X=x_2$.}
\label{Figbif}
\end{figure}

Modal manifolds can merge or bifurcate as $x$ varies. Interestingly,
though, the merges or bifurcations do not necessarily occur at points
of contact between two modal manifolds. See Figure~\ref{Figbif} for
an example with $d=1$. Shown is a modal curve (manifold
with $d=1$), starting at $x=0$ and stopping at about $x=0.35$, which
leaves a gap between itself and the neighboring modal curve. We take
a closer look at the joint density contours, in panel (c), and inspect
the conditional density along four slices $X=x_1,\ldots,x_4$, in
panel (d). We see that after $X=x_2$, the conditional density becomes
unimodal and the first (left) mode disappears, as it slides into a
saddle point.

A remark about the uniqueness of the modal manifold collection
$\mathcal{S}$ in \eqref{eqS1}: this factorization is unique if the
second derivative $p_{yy}(x,y)$ is uniformly bounded away from zero.
This will later become one of our assumptions [assumption (A3)] in the
theoretical analysis of Section~\ref{secasymptotics}. Note that in
the left panel of Figure~\ref{Figthm1}, the collection $\mathcal{S}$
is uniquely defined, while this is not the case in the right panel
(at the points of intersection between curves, the density $p$ has
vanishing second derivatives with respect to $y$).

Lastly, the population quantities defined above all
have sample analogs. For the estimate {$\hat{M}_n(x)$},
we define
%
\begin{equation}
\label{eqES1} \hat{\mathcal{S}}_n = \bigl\{ (x,y ):y\in
\hat{M}_n(x), x\in\R\bigr\} = \hat{S}_1\cup\cdots\cup
\hat{S}_{\hat{K}},
\end{equation}
where each {$\hat{S}_j$} is a connected manifold, and
{$\hat{K}$} is the total number.
We also define {$\hat{m}_j(x)$} in a similar way for
{$j=1,\ldots,\hat{K}$}. Thus, we can write
%
\begin{equation}
\label{eqEMFdef} \hat{M}_n(x) = \bigl\{\hat{m}_1(x),
\ldots,\hat{m}_{\hat{K}}(x) \bigr\}.
\end{equation}
%
In practice, determining the manifold
memberships and the total number of manifolds {$\hat{K}$} is not
trivial. In principle, the sample manifolds
{$\hat{S}_1,\ldots, \hat{S}_{\hat{K}}$} are well-defined in
terms of the sample estimate {$\hat{M}_n(x)$}; but
even with a perfectly convergent mean-shift algorithm, we would need
to run mean-shift iterations at every input $x$ in the domain $\K$ to
determine these manifold components. Clearly, this is not an
implementable strategy. Thus, from the output of the mean-shift
algorithm over a finite mesh, we usually employ
some type of simple post-processing
technique to determine connectivity of the outputs, and hence the
sample manifolds. This is discussed further in Section~\ref{secmixture}.

\section{Asymptotic error analysis}\label{secasymptotics}

In this section, we present asymptotic results about the convergence of
the estimated modal regression set {$\hat{M}_n(x)$} to the
underlying modal set $M(x)$.
Let $\mathbf{BC}^k(C)$ denote the collection of $k$ times continuously
differentiable functions with all partial derivatives bounded in
absolute value by $C$. (The domain of these functions should be
clear from the context.) Given a kernel function
$K: \R\rightarrow\R$, denote the collection of functions
\[
\mathcal{K} = \biggl\{v\mapsto K^{(\alpha)} \biggl(\frac
{z-v}{h}
\biggr): z\in\R, h>0, \alpha=0,1,2 \biggr\},
\]
where $K^{(\alpha)}$ denotes the $\alpha$th order derivative of $K$.

Our assumptions are as follows.
\begin{longlist}[(K2)]
\item[(A1)] The joint density $p\in\mathbf{BC}^4(C_p)$ for some
$C_p>0$.
\item[(A2)] The collection of modal manifolds $\mathcal{S}$ can be
factorized into $\mathcal{S} = S_1\cup\cdots\cup S_K$,
where each $S_j$ is a connected curve
that admits a parametrization $S_j = \{(x,m_j(x)):x\in A_j\}$
for some $m_j(x)$, and $A_1,\ldots, A_K$ form an open cover
for the support $\K$ of $X$.
\item[(A3)] There exists $\lambda_2>0$ such that for any
$(x,y)\in\K\times\mathbb{K}$ with $p_y(x,y)=0$,
$\llvert p_{yy}(x,y)\rrvert>\lambda_2$.
%
\item[(K1)] The kernel function $K\in\mathbf{BC}^2(C_K)$ and
satisfies
\[
\int_\R\bigl(K^{(\alpha)}\bigr)^2(z) \,dz <
\infty, \qquad\int_\R z^2K^{(\alpha)}(z)
\,dz<\infty,
\]
for $\alpha=0,1,2$.
\item[(K2)] The collection $\mathcal{K}$ is a VC-type
class, that is, there exists $A,v>0$ such that for $0<\varepsilon<1$,
\[
\sup_Q N \bigl(\mathcal{K}, L_2(Q),
C_K\varepsilon\bigr) \leq\biggl(\frac{A}{\varepsilon}
\biggr)^v,
\]
where $N(T, d,\varepsilon)$ is the $\varepsilon$-covering number for a
semimetric space $(T,d)$ and $Q$ is any probability measure.
\end{longlist}

Assumption (A1) is an ordinary smoothness condition; we need
fourth derivatives since we need to bound the bias of second
derivatives. The assumption (A2) is to make sure the collection of all
local modes can be represented as finite collection of manifolds.
(A3) is a sharpness requirement for all critical points (local modes
and minimums); and
it excludes the case that the modal manifolds
bifurcate or merge, that is, it excludes cases such as the right panel of
Figure~\ref{Figthm1}.
Similar
conditions appear in \citet{romano1988weak,Chen2014GMRE} for
estimating density modes.
Assumption (K1) is assumed for the
kernel density estimator to have the usual rates for its bias and
variance. (K2) is for the uniform bounds on the kernel density
estimator; this condition can be found in
\citet{gine2002rates,einmahl2005uniform,Chen2014ridge}.
We study three types of error metrics for regression modes:
pointwise, uniform and mean integrated squared errors. We defer all
proofs to the supplementary material [\citet{Chen2015Modalsupp}].

First, we study the pointwise case. Recall that {$\hat{p}_n$} is
the KDE in \eqref{eqKDE} of the joint density based on $n$ samples,
under the kernel $K$, and {$\hat{M}_n(x)$} is the estimated
modal regression set in \eqref{eqMhat2} at a point $x$. Our
pointwise analysis considers
\[
\Delta_n(x) = \Haus\bigl(\hat{M}_n(x),M(x) \bigr),
\label{eqUM0}
\]
the Hausdorff distance between {$\hat{M}_n(x)$} and $M(x)$, at a
point $x$. For our first result, we define the quantities:
\begin{eqnarray*}
\llVert\hat{p}_n-p\rrVert^{(0)}_{\infty} &=& \sup
_{x,y} \bigl\llVert\hat{p}(x,y)-p(x,y)\bigr\rrVert,
\\
\llVert\hat{p}_n-p\rrVert^{(1)}_{\infty} &=& \sup
_{x,y}\bigl\llVert\hat{p}_y(x,y)-p_y(x,y)
\bigr\rrVert,
\\
\llVert\hat{p}_n-p\rrVert^{(2)}_{\infty} &=& \sup
_{x,y}\bigl\llVert\hat{p}_{yy}(x,y)-p_{yy}(x,y)
\bigr\rrVert,
\\
\llVert\hat{p}_n-p\rrVert^{*}_{\infty,2} &=& \max
\bigl\{\llVert\hat{p}_n-p\rrVert^{(0)}_{\infty},
\llVert\hat{p}_n-p\rrVert^{(1)}_{\infty}, \llVert
\hat{p}_n-p\rrVert^{(2)}_{\infty} \bigr\}.
\end{eqnarray*}

%
\begin{teo}[(Pointwise error rate)]\label{teoW2}
Assume \textup{(A1)--(A3)} and \textup{(K1)--(K2)}.
Define a stochastic process $A_n(x)$ by
\[
A_n(x) =\cases{ \displaystyle\frac{1}{\Delta_n(x)}\Bigl\llvert\Delta_n(x)-
\max_{z \in M(x)} \bigl\{\bigl\llvert p^{-1}_{yy}(x,z)
\bigr\rrvert\bigl\llvert\hat{p}_{y,n}(x,z)\bigr\rrvert\bigr\}\Bigr
\rrvert,
\vspace*{3pt}\cr
\hspace*{33pt} \mbox{if $\Delta_n(x)>0$,}
\vspace*{3pt}\cr
0, \qquad\mbox{if $\Delta_n(x)=0$}.}
\]
Then, when
\[
\llVert\hat{p}_n-p\rrVert^{*}_{\infty,2} = \max
\bigl\{\llVert\hat{p}_n-p\rrVert^{(0)}_{\infty},
\llVert\hat{p}_n-p\rrVert^{(1)}_{\infty}, \llVert
\hat{p}_n-p\rrVert^{(2)}_{\infty} \bigr\}
\]
is sufficiently small, we have
\[
\sup_{x \in D} A_n(x) = O_\P\bigl(\llVert
\hat{p}_n-p\rrVert^{*}_{\infty,2}\bigr).
\]
Moreover, at any fixed $x \in\K$,
when {$\frac{nh^{d+5}}{\log n}\rightarrow\infty$} and
$h\rightarrow0$,
\[
\Delta_n(x) = O\bigl(h^2\bigr)+O_\P\biggl(
\sqrt{\frac{1}{nh^{d+3}}} \biggr).
\]
\end{teo}

The proof is in the supplementary material [\citet{Chen2015Modalsupp}].
This shows that if the curvature of the joint density function
along $y$ is bounded away from $0$, then the error can be
approximated by the error of {$\hat{p}_{y,n}(x,z)$} after
scaling. The rate of convergence follows from the fact that
{$\hat{p}_{y,n}(x,z)$} is converging to $0$ at the same rate.
Note that as $z$ is a conditional mode, the partial derivative
of the true density is~$0$.
We defined $A_n(x)$ as above since
$\Delta_n(x)=0$ implies $\max_{z \in M(x)}\llvert\hat
{p}_{y,n}(x,z)\rrvert=0$,
so that the ratio would be ill-defined if $\Delta_n(x)=0$.
Also, the constraints on $h$ in the second assertion
({$\frac{nh^{d+5}}{\log n}\rightarrow\infty$ and
$h\rightarrow0$})
are to ensure $\llVert\hat{p}_n-p\rrVert^{*}_{\infty,2}=o_\P(1)$.

For our next result, we define the uniform error
\[
\Delta_n = \sup_{x \in\K} \Delta_n(x) =
\sup_{x \in\K} \Haus\bigl(\hat{M}_n(x),M(x) \bigr).
\label{eqUM1}
\]
This is an $\ell_{\infty}$ type error for estimating regression
modes (and is also closely linked to confidence sets; see Section~\ref
{secconfidence}).

%
\begin{teo}[(Uniform error rate)]\label{teoW3}
Assume \textup{(A1)--(A3)} and \textup{(K1)--(K2)}. Then as
{$\frac{nh^{d+5}}{\log n}\rightarrow\infty$} and
$h\rightarrow0$,
\[
\Delta_n = O\bigl(h^2\bigr)+O_\P\biggl(
\sqrt{\frac{\log n}{nh^{d+3}}} \biggr).
\]
\end{teo}

The proof is in the supplementary material [\citet{Chen2015Modalsupp}].
Compared to the pointwise error rate in Theorem \ref{teoW2}, we
have an additional {$\sqrt{\log n}$} factor in the second term.
One can view this as the price we need to pay for getting an uniform
bound over all points. See \citet{gine2002rates,einmahl2005uniform} for similar
findings in density estimation.

The last error metric we consider is the mean integrated squared error
(MISE), defined as
\[
\MISE(\hat{M}_n ) = \mathbb{E} \biggl(\int_{x\in\K}
\Delta^2_n(x) \,dx \biggr). \label{eqMISEdef}
\]
Note that the MISE is a nonrandom quantity, unlike first two error
metrics considered.

%
\begin{teo}[(MISE rate)]\label{teoW4}
Assume \textup{(A1)--(A3)} and \textup{(K1)--(K2)}. Then as $\frac{nh^{d+5}}{\log n}\rightarrow
\infty$ and $ h\rightarrow0$,
\[
\MISE(\hat{M}_n ) = O\bigl(h^4\bigr)+O \biggl(
\frac
{1}{nh^{d+3}} \biggr).
\]
\end{teo}

The proof is in the supplementary material [\citet{Chen2015Modalsupp}].
If we instead focus on estimating the regression modes of
the smoothed joint density {$\tilde{p}(x,y) =
\mathbb{E} (\hat{p}_n(x,y) )$}, then we obtain much
faster convergence rates.
Let {$\tilde{M}(x) = \mathbb{E}(\hat{M}_n(x))$} be
the smoothed regression modes at $x\in\K$. Analogously, define
\begin{eqnarray*}
\tilde{\Delta}_n(x) &=& \Haus\bigl(\hat{M}_n(x),
\tilde{M}(x) \bigr),
\\
\tilde{\Delta}_n &=& \sup_{x \in\K} \tilde{
\Delta}_n(x),
\\
\tilde{\MISE} (\hat{M}_n ) & =& \mathbb{E} \biggl(\int
_{x\in\K}\tilde{\Delta}^2_n(x) \,dx \biggr).
\end{eqnarray*}

%
\begin{cor}[(Error rates for smoothed conditional modes)]\label{corW5}
Assume \textup{(A1)--(A3)} and \textup{(K1)--(K2)}. Then as
$\frac{nh^{d+5}}{\log n}\rightarrow
\infty$ and $h\rightarrow0$,
\begin{eqnarray*}
\sqrt{nh^{d+3}}\sup_{x \in D}\Bigl\llvert\tilde{
\Delta}_n(x) -\max_{z\in\tilde{M}(x)}\bigl\{\tilde{p}^{-1}_{yy}(x,z)
\hat{p}_{y,n}(x,z) \bigr\}\Bigr\rrvert &=& O_\P(
\varepsilon_{n,2}),
\\
\tilde{\Delta}_n(x) &=&O_\P\biggl(\sqrt{
\frac{1}{nh^{d+3}}} \biggr),
\\
\tilde{\Delta}_n &=& O_\P\biggl(\sqrt{\frac{\log
n}{nh^{d+3}}}
\biggr),
\\
\tilde{\MISE} (\hat{M}_n ) &=& O \biggl(\frac
{1}{nh^{d+3}}
\biggr),
\end{eqnarray*}
where
\[
\varepsilon_{n,2} = \sup_{x,y}\bigl\llvert\hat{p}_{yy,n}(x,y)-\tilde
{p}_{yy}(x,y)\bigr\rrvert
=\sup_{x,y}\bigl\llvert\hat{p}_{yy,n}(x,y)-\mathbb{E}\bigl(\hat
{p}_{yy,n}(x,y)\bigr)\bigr\rrvert.
\]
\end{cor}


\section{Confidence sets}\label{secconfidence}

In an idealized setting, we could define a confidence set at $x$ by
\[
\hat{C}^0_n(x) = \hat{M}_n(x)\oplus
\delta_{n,1-\alpha}(x),
\]
where
\[
\P\bigl(\Delta_n(x) > \delta_{n,1-\alpha}(x) \bigr) = \alpha.
\label{eqdeltadef}
\]
By construction, we have
{$\P(M(x) \in\hat{C}^0_n(x)) = 1-\alpha$}.
Of course, the distribution of $\Delta_n(x)$ is unknown, but we
can use the bootstrap~[\citet{Efron1979}] to estimate
$\delta_{n,1-\alpha}(x)$.

Given the observed data samples $(X_1,Y_1),\ldots,(X_n,Y_n)$, we
denote a bootstrap sample as $(X_1^*, Y_1^*),
\ldots,(X^*_n,Y^*_n)$. Let $\hat{M}^*_n(x)$ be the
estimated regression modes based on this bootstrap sample, and
\[
\hat{\Delta}_n^*(x) = \Haus\bigl(\hat{M}^*_n(x),
\hat{M}_n(x) \bigr). \label{eqEdeltadef}
\]
We repeat the bootstrap sampling $B$ times to get
$\hat{\Delta}_{1, n}^*(x),\ldots,\hat{\Delta}_{B, n}^*(x)$.
Define $\hat\delta_{n,1-\alpha}(x)$ by
\[
\frac{1}{B} \sum_{j=1}^B I \bigl(
\hat{\Delta}_{j, n}^*(x) > \hat\delta_{n,1-\alpha}(x) \bigr) = \alpha.
\]
Our confidence set for $M(x)$ is then given by
\[
\hat{C}_n(x) = \hat{M}_n(x)\oplus\hat{
\delta}_{n,1-\alpha}(x). \label{eqPC1}
\]
Note that this is a pointwise confidence set, at $x \in\K$.

Alternatively, we can use $\Delta_n = \sup_{x \in\K} \Delta_n(x)$
to build a uniform confidence set. Define $\delta_{n,1-\alpha}$ by
\[
\P\bigl(M(x)\subseteq\hat{M}_n(x)\oplus\delta_{n,1-\alpha},
\forall x \in\K\bigr)= 1-\alpha. \label{eqCI1}
\]
As above, we can use bootstrap sampling to form an estimate
{$\hat{\delta}_{n,1-\alpha}$}, based on the quantiles of the
bootstrapped uniform error metric
\[
\hat{\Delta}_n^* = \sup_{x\in\K} \Haus\bigl(
\hat{M}^*_n(x),\hat{M}_n(x) \bigr). \label{eqCI2}
\]
Our uniform confidence set is then
\[
\hat{C}_n = \bigl\{(x,y): x\in\K, y\in\hat{M}_n(x)\oplus
\hat{\delta}_{n,1-\alpha} \bigr\}. \label{eqCI3}
\]

\begin{figure}

\includegraphics{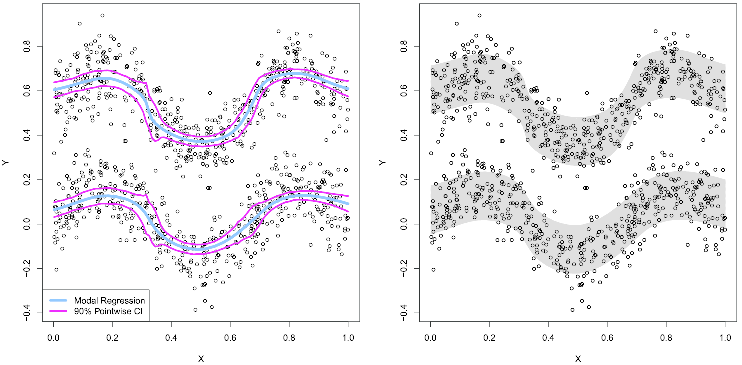}

\caption{An example with pointwise (left) and uniform (right) confidence
sets. The significance level is 90\%.}
\label{Figconf}
\end{figure}

In practice, there are many possible flavors of the bootstrap that are
applicable here. This includes the ordinary (nonparametric)
bootstrap, the smoothed bootstrap and the residual bootstrap. See
Figure~\ref{Figconf} for an example with the ordinary bootstrap.

We focus on the asymptotic coverage of uniform
confidence sets built with the ordinary bootstrap. We
consider coverage of the smoothed regression mode set
{$\tilde{M}(x)$} (to avoid issues of bias), and we employ tools
developed in \citet{Chen2014ridge},
Chernozhukov, Chetverikov and Kato (\citeyear{chernozhukov2014anti,chernozhukov2014gaussian}).

%

Consider a function space $\mathcal{F}$ defined as
\begin{eqnarray}\label{eqBT2}
\mathcal{F} &=& \biggl\{ (u,v) \mapsto f_{x,y}(u,v): f_{x,y}(u,v)
= \tilde{p}^{-1}_{yy}(x,y)
\nonumber\\[-8pt]\\[-8pt]\nonumber
&&{}\times K \biggl(\frac{\llVert x-u\rrVert}{h} \biggr) K^{(1)} \biggl(\frac{y-v}{h}
\biggr), x \in\K, y\in\tilde{M}(x) \biggr\},
\end{eqnarray}
and let $\mathbb{B}$ be a Gaussian process defined on
$\mathcal{F}$ such that
\begin{eqnarray}\label{eqBT1}
&& \Cov\bigl(\mathbb{B}(f_1),\mathbb{B}(f_2)\bigr)
\nonumber\\[-8pt]\\[-8pt]\nonumber
&&\qquad = \mathbb{E} \bigl(f_1(X_i,Y_i)f_2(X_i,Y_i)
\bigr)- \mathbb{E} \bigl(f_1(X_i,Y_i)
\bigr)\mathbb{E} \bigl(f_2(X_i,Y_i) \bigr),
\end{eqnarray}
for all $f_1,f_2\in\mathcal{F}$.

%
\begin{teo}
\label{teoBT2-1}
Assume \textup{(A1)--(A3)} and \textup{(K1)--(K2)}.
Define the random variable
$\mathbf{B} = \frac{1}{\sqrt{h^{d+3}}}\sup_{f\in\mathcal
{F}}\llvert\mathbb{B}(f)\rrvert$.
Then as $\frac{nh^{d+5}}{\log n}\rightarrow\infty$, $h\rightarrow0$,
\[
\sup_{t \geq0} \bigl\llvert\mathbb{P} \bigl(\sqrt{nh^{d+3}}
\tilde{\Delta}_n<t \bigr)- \mathbb{P} (\mathbf{B}<t ) \bigr\rrvert=O
\biggl( \biggl(\frac{\log^7 n}{nh^{d+3}} \biggr)^{1/8} \biggr).
\]
\end{teo}

The proof is in the supplementary material [\citet{Chen2015Modalsupp}].
This theorem shows that the smoothed uniform discrepancy
{$\tilde{\Delta}_n$} is distributed asymptotically as the
supremum of a Gaussian process. In fact, it can be shown that the two
random variables {$\tilde{\Delta}_n$} and $\mathbf{B}$ are
coupled by
\[
\bigl\llvert\sqrt{nh^{d+3}}\tilde{\Delta}_n - \mathbf{B}
\bigr\rrvert= O_\P\biggl( \biggl(\frac{\log^7 n}{nh^{d+3}}
\biggr)^{1/8} \biggr).
\]

Now we turn to the limiting behavior for the bootstrap estimate.
Let $\mathcal{D}_n = \{(X_1,Y_1),\ldots,(X_n,Y_n)\}$ be the observed
data and
denote the bootstrap estimate by
\[
\hat{\Delta}^*_n = \sup_{x\in\K} \Haus\bigl(
\hat{M}^*_n(x),\hat{M}_n(x) \bigr),
\]
where $\hat{M}^*_n(x)$ is the bootstrap regression mode set at $x$.

%
\begin{teo}[(Bootstrap consistency)]
\label{teoBT2-2}
Assume conditions \textup{(A1)--(A3)} and \mbox{\textup{(K1)--(K2)}}.
Also assume that
$\frac{nh^{d+5}}{\log n}\rightarrow\infty, h\rightarrow0$.
Define
\[
\mathbf{B} = \frac{1}{\sqrt{h^{d+3}}}\sup_{f\in\mathcal
{F}}\bigl\llvert\mathbb{B}(f)\bigr\rrvert.
\]
There exists
$\mathcal{X}_n$ such that
$\mathbb{P}(\mathcal{X}_n)\geq1-O(\frac{1}{n})$ and,
for all
$\mathcal{D}_n\in\mathcal{X}_n$,
\[
\sup_{t \geq0} \bigl\llvert\mathbb{P} \bigl(\sqrt{nh^{d+3}}
\hat{\Delta}^*_n<t \mid\mathcal{D}_n \bigr)- \mathbb{P} (
\mathbf{B}<t ) \bigr\rrvert=O_\P\biggl( \biggl(\frac{\log^7 n}{nh^{d+3}}
\biggr)^{1/8} \biggr).
\]
\end{teo}

The proof is in the supplementary material [\citet{Chen2015Modalsupp}].
Theorem~\ref{teoBT2-2} shows that
the limiting distribution for the bootstrap estimate
{$\hat{\Delta}^*_n$} is the same
as the limiting distribution of {$\tilde{\Delta}_n$} (recall
Theorem \ref{teoBT2-1}) with high probability.
Using Theorems~\ref{teoBT2-1}~and~\ref{teoBT2-2}, we
conclude the following.

%
\begin{cor}[(Uniform confidence sets)]
\label{corBT2-3}
Assume\vspace*{1pt} conditions \textup{(A1)--(A3)} and \textup{(K1)--(K2)}. Then as $\frac{nh^{d+5}}{\log
n}\rightarrow\infty$ and $h\rightarrow0$,
\[
\mathbb{P} \bigl( \tilde{M}(x)\subseteq\hat{M}_n(x)\oplus\hat{
\delta}_{n,1-\alpha}, \forall x\in\K\bigr)= 1-\alpha+O \biggl( \biggl(
\frac{\log^7 n}{nh^{d+3}} \biggr)^{1/8} \biggr).
\]
\end{cor}

\section{Prediction sets}\label{secprediction}

Modal regression can be also used to construct prediction sets.
Define
\begin{eqnarray*}
\varepsilon_{1-\alpha}(x) &=&\inf\bigl\{\varepsilon\geq0: \P\bigl
(d\bigl(Y,M(X)
\bigr)>\varepsilon\mid X=x \bigr)\leq\alpha\bigr\},
\\
\varepsilon_{1-\alpha} &=&\inf\bigl\{\varepsilon\geq0: \P\bigl(d\bigl(Y,M(X)
\bigr)>\varepsilon\bigr)\leq\alpha\bigr\}.
\end{eqnarray*}
Recall that $d(x,A)=\inf_{y\in A}\llvert x-y\rrvert$ for a
point $x$ and a set
$A$. Then
\begin{eqnarray*}
\mathcal{P}_{1-\alpha}(x) &=& M(x)\oplus\varepsilon_{1-\alpha
}(x)\subseteq\R,
\\
\mathcal{P}_{1-\alpha} &=& \bigl\{(x,y): x \in\K, y\in M(x)\oplus
\varepsilon_{1-\alpha} \bigr\}\subseteq\K\times\R
\end{eqnarray*}
are pointwise and uniform prediction sets, respectively, at the
population level, because\vspace*{-4pt}
\begin{eqnarray*}
\P\bigl(Y\in{\mathcal P}_{1-\alpha}(x) \mid X=x \bigr) &\geq&1-\alpha,
\\
\P(Y\in{\mathcal P}_{1-\alpha}) &\geq&1-\alpha.
\end{eqnarray*}

At the sample level, we use a KDE of the conditional density
{$\hat{p}_n(y\mid x) = \hat{p}_n(x,y)/\hat{p}_n(x)$}, and estimate
$\varepsilon_{1-\alpha}(x)$ via
\[
\hat{\varepsilon}_{1-\alpha}(x) = \inf\biggl\{\varepsilon\geq0:\int
_{\hat{M}_n(x)\oplus\varepsilon} \hat{p}_n(y\mid x) \,dy\geq1-\alpha
\biggr\}.
\label{eqwEPPS}
\]
An estimated pointwise prediction set is then
\[
\hat{\mathcal{P}}_{1-\alpha}(x) = \hat{M}_n(x)\oplus\hat{
\varepsilon}_{1-\alpha}(x). \label{eqEPPS}
\]
This has the proper pointwise coverage with respect to
samples drawn according to {$\hat{p}_n(y\mid x)$}, so in an
asymptotic regime in which
{$\hat{p}_n(y\mid x) \rightarrow p_n(y\mid x)$}, it will have the
correct coverage with respect to the population distribution, as well.

Similarly, we can define
%
\begin{equation}
\hat{\varepsilon}_{1-\alpha} = \Quantile\bigl( \bigl\{d \bigl(Y_i,
\hat{M}_n(X_i) \bigr): i=1,\ldots,n \bigr\}, 1-\alpha
\bigr), \label{eqwEPS}
\end{equation}
the $(1-\alpha)$ quantile of
{$d(Y_i, \hat{M}_n(X_i))$}, $i=1,\ldots, n$, and then the estimated
uniform prediction set is
%
\begin{equation}
\hat{\mathcal{P}}_{1-\alpha}= \bigl\{(x,y): x\in\K, y \in
\hat{M}_n(x)\oplus\hat{\varepsilon}_{1-\alpha} \bigr\}.
\label{eqEPS}
\end{equation}
The estimated uniform prediction set has
proper coverage with respect to the empirical distribution, and so
certain conditions, it will have valid limiting population coverage.\vspace*{-1pt}

\subsection{Bandwidth selection}\label{secbandwidth}

Prediction sets can be used to select the smoothing
bandwidth of the underlying KDE, as we describe here. We focus on
uniform prediction sets, and we will use a
subscript $h$ throughout to denote the dependence on the smoothing
bandwidth. From its definition in \eqref{eqEPS}, we can see that the
volume (Lebesgue measure) of the estimated uniform prediction set is\vspace*{-1pt}
\[
\Vol(\hat{\mathcal{P}}_{1-\alpha,h} ) = \hat{\varepsilon}_{1-\alpha,h}
\int
_{x\in\K} \hat{K}_h(x) \,dx, \label{eqVh}
\]
where {$\hat{K}_h(x)$} is the number of estimated local modes at
$X=x$, and {$\hat{\varepsilon}_{1-\alpha,h}$} is as defined in
\eqref{eqwEPS}.
Roughly speaking, when $h$ is small, $\hat{\varepsilon}_{1-\alpha,h}$ is
also small, but the number of estimated manifolds is large; on the
other hand, when $h$ is large, $\hat{\varepsilon}_{1-\alpha,h}$ is large,
but the number of estimated manifolds is small. This is like the
bias-variance trade-off: small $h$ corresponds to less bias
({$\hat{\varepsilon}_{1-\alpha,h}$}) but higher variance (number of
estimated manifolds).

\begin{figure}

\includegraphics{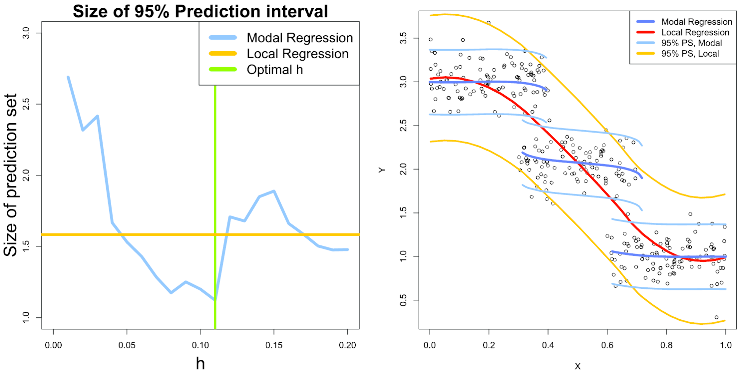}

\caption{An example of bandwidth selection based on the size of the
prediction sets.}
\label{Figh}
\end{figure}

Our proposal is to select $h$ by\vspace*{-2pt}
\[
h^* = \mathop{\operatorname{argmin}}_{h \geq0} \Vol(\hat{
\mathcal{P}}_{1-\alpha,h} ).
\]
Figure~\ref{Figh} gives an example this rule when $\alpha=0.05$,
that is, when minimizing the size of the estimated 95\% uniform
prediction set.
Here, we actually use cross-validation to obtain the size of the
prediction set; namely, we use the training set to estimate the modal
manifolds and then use the validation set to estimate the width of
prediction set. This helps us to avoid overfitting.
As can be seen, there is a clear trade-off in the
size of the prediction set versus $h$ in the left plot.
The optimal value $h^*=0.07$ is marked by a vertical line,
and the right plot displays the corresponding modal regression
estimate and uniform prediction set on the data samples.

%

In the same plot, we also display a local regression estimate and its
corresponding 95\% uniform prediction set. We can see that the
prediction set from the local regression method is much larger than
that from modal regression. (To even the comparison, the bandwidth for
the local linear smoother was also chosen to minimize the size of the
prediction set.)
This illustrates a major strength of the modal regression method:
because it is not constrained to modeling conditional mean structure,
it can produce smaller prediction sets than the usual regression
methods when the conditional mean fails to capture the main structure
in the data. We investigate this claim theoretically, next.

\subsection{Theory on the size of prediction sets}

We will show that, at the population level,
prediction sets from modal regression can be smaller than those based
on the underlying regression function $\mu(x)=\mathbb{E}(Y\mid X=x)$.
Defining
\begin{eqnarray*}
\eta_{1-\alpha}(x) &=&\inf\bigl\{\eta\geq0: \P\bigl(d\bigl(Y,\mu
(X)\bigr)>
\eta\mid X=x \bigr)\leq\alpha\bigr\},
\\
\eta_{1-\alpha} &=&\inf\bigl\{\eta\geq0: \P\bigl(d\bigl(Y,\mu(X)\bigr)>
\eta\bigr)\leq\alpha\bigr\},
\end{eqnarray*}
pointwise and uniform prediction sets based on the regression function
are
\begin{eqnarray*}
\mathcal{R}_{1-\alpha}(x) & =& \mu(x)\oplus\eta_{1-\alpha
}(x)\subseteq\R,
\\
\mathcal{R}_{1-\alpha} &=& \bigl\{\bigl(x,\mu(x)\oplus\eta_{1-\alpha}
\bigr): x\in\K\bigr\}\subseteq\K\times\R,
\end{eqnarray*}
respectively.

For a pointwise prediction set $A(x)$, we write
$\length(A(x))$ for its Lebesgue measure on $\R$; note that
in the case of modal regression, this is somewhat of an abuse of
notation because the Lebesgue measure of $A(x)$ can be a sum of
interval lengths. For a uniform prediction set $A$, we write
$\Vol(A)$ for its Lebesgue measure on $\K\times\R$.

We consider the following assumption.
\begin{longlist}
\item[(GM)] The conditional density satisfies
\[
p(y\mid x) = \sum_{j=1}^{K(x)}
\pi_j(x) \phi\bigl(y; \mu_j(x), \sigma_j^2(x)
\bigr)
\]
with $\mu_1(x)<\mu_2(x)<\cdots<\mu_{K(x)}(x)$ by convention, and
$\phi( \cdot; \mu, \sigma^2)$ denoting the Gaussian density
function with mean $\mu$ and variance $\sigma^2$.
\end{longlist}
The assumption that the conditional density can be written as a
mixture of Gaussians is only used for the next result. It is
important to note that this is an assumption made about the
population density, and does not reflect modeling choices made in the
sample. Indeed, recall, we are comparing prediction sets based on the
modal set $M(x)$ and the regression function $\mu(x)$, both of which
use true population information.

Before stating the result, we must define several quantities.
Define the minimal separation between mixture centers
\[
\Delta_{\min}(x) = \min\bigl\{\bigl\llvert\mu_i(x)-
\mu_j(x)\bigr\rrvert: i\neq j\bigr\}
\]
and
\begin{eqnarray*}
\sigma^2_{\max}(x) &=&\max_{j=1,\ldots, K(x)}
\sigma^2_j(x),
\\
\pi_{\max}(x) &=&\max_{j=1,\ldots, K(x)} \pi_j(x),\qquad
\pi_{\min}(x) =\min_{j=1,\ldots, K(x)} \pi_j(x).
\end{eqnarray*}
Also define
\[
\Delta_{\min} = \inf_{x \in\K} \Delta_{\min}(x),
\qquad\sigma^2_{\max}=\sup_{x \in\K}
\sigma^2_{\max}(x),
\]
and
\[
\pi_{\max}=\sup_{x\in\K} \pi_{\max}(x), \qquad
\pi_{\min}=\inf_{x\in\K} \pi_{\min}(x),
\]
and
\[
\overline{K}= \frac{\int_{x \in\K} K(x) \,dx}{\int_{x \in\K}
\,dx},\qquad K_{\min} = \inf
_{x \in\K} K(x),\qquad K_{\max} = \inf_{x\in\K}
K(x).
\]

%
\begin{teo}[(Size of prediction sets)]\label{teoPPS}
Assume \textup{(GM)}. Let $\alpha<0.1$ and assume that
$\pi_1(x),\pi_{K(x)}(x)>\alpha$.
If
\begin{eqnarray*}
\frac{\Delta_{\min}(x)}{\sigma_{\max}(x)}&>& \max\biggl\{1.1\cdot\frac
{K(x)}{K(x)-1} z_{1-\alpha/2},
\\
&&{} \sqrt{6.4\vee2\log\bigl(4\bigl(K(x)\vee3 -1 \bigr) \bigr)+ 2\log\biggl(
\frac{\pi_{\max}(x)}{\pi_{\min}(x)} \biggr)} \biggr\},
\end{eqnarray*}
where $z_{\alpha}$ is the upper $\alpha$-quantile value of a standard
normal distribution
and $A\vee B = \max\{A,B\}$, then
\[
\length\bigl(\mathcal{P}_{1-\alpha}(x) \bigr)< \length\bigl(
\mathcal{R}_{1-\alpha}(x) \bigr).
\]
Moreover, if
\begin{eqnarray*}
\frac{\Delta_{\min}}{\sigma_{\max}}&>&\max\biggl\{ 1.1\cdot\biggl(\frac
{2\overline{K}}{K_{\min}-1} \biggr)
z_{1-\alpha/2},
\\
&&{}\sqrt{6.4\vee2\log\bigl(4(K_{\max}\vee3 -1 ) \bigr)+ 2\log\biggl(
\frac{\pi_{\max}}{\pi_{\min}} \biggr)} \biggr\},
\end{eqnarray*}
then
\[
\Vol(\mathcal{P}_{1-\alpha})< \Vol(\mathcal{R}_{1-\alpha}).
\]
\end{teo}

The proof is in the supplementary material [\citet{Chen2015Modalsupp}].
In words, the theorem shows that when the signal-to-noise ratio is
sufficiently large, the modal-based prediction set is smaller than the
usual regression-based prediction set.


\section{Comparison to mixture regression}\label{secmixture}

Mixture regression is similar to modal
regression. The literature on mixture regression, also known as
mixture of experts modeling, is vast; see, for example,
\citet
{Jacobs1991,Jiang1999,Bishop2006,viele2002modeling,Khalili2007jasa,hunter2012semiparametric,Huang2012jasa,Huang2013jasa}.
In mixture regression, we assume that the conditional density function
takes the form
\[
p(y\mid x) = \sum_{j=1}^{K(x)}
\pi_j(x) \phi_j\bigl(y; \mu_j(x),
\sigma_j^2(x)\bigr),
\]
where each {$\phi_j(y; \mu_j(x), \sigma_j^2(x))$} is a density
function, parametrized by a mean $\mu_j(x)$ and variance
{$\sigma_j^2(x)$}. The\vspace*{2pt} simplest and most common usage of
mixture regression makes the following assumptions:
\begin{longlist}[(MR5)]
\item[(MR1)] $K(x)=K$,
\item[(MR2)] $\pi_j(x) = \pi_j$ for each $j$,\vspace*{1pt}
\item[(MR3)] $\mu_j(x) = \beta_j^T x$ for each $j$,\vspace*{1pt}
\item[(MR4)] $\sigma^2_j(x) = \sigma_j^2$ for each $j$, and\vspace*{1pt}
\item[(MR5)] $\phi_j(x)$ is Gaussian for each $j$.
\end{longlist}
This is called linear mixture regression
[\citet{viele2002modeling,chaganty2013spectral}]. Many authors have
considered relaxing some subset of the above assumptions, but as far
we can tell, no work has been proposed to effectively relax
all of (MR1)--(MR5).

Modal regression is a fairly simple tool that achieves a similar
goal to mixture regression models, and uses fewer assumptions.
Mixture regression is inherently a model-based method,
stemming from a model for the joint density $p(y\mid x)$; modal
regression hunts directly for conditional modes, which can be
estimated without a model for $p(y\mid x)$. Another important difference:
the number of mixture components $K$ in the mixture regression model
plays a key role, and estimating $K$ is quite difficult;
in modal regression we do not need to estimate anything of this sort
(e.g., we do not specify a number of modal manifolds). Instead, the
flexibility of the estimated modal regression set is driven by the
bandwidth parameter $h$ of the KDE, which can be tuned by
inspecting the size of prediction sets, as described in Section~\ref
{secbandwidth}.
Table~\ref{tabMM} summarizes the comparison between mixture-based
and mode-based methods.

%
\begin{table}[b]
\tabcolsep=0pt
\caption{Comparison for methods based on mixtures versus modes}
\label{tabMM}
\begin{tabular*}{\tablewidth}{@{\extracolsep{\fill}}@{}lcc@{}}
\hline
& \textbf{Mixture-based} & \textbf{Mode-based}\\
\hline
Density estimation & Gaussian mixture & Kernel density estimate \\
Clustering & $K$-means & Mean-shift clustering \\
Regression & Mixture regression & Modal regression \\
Algorithm & EM & Mean-shift \\
Complexity parameter & $K$ (number of components) & $h$ (smoothing bandwidth) \\
Type & Parametric model & Nonparametric model \\
\hline
\end{tabular*}
\end{table}

\begin{figure}[t]

\includegraphics{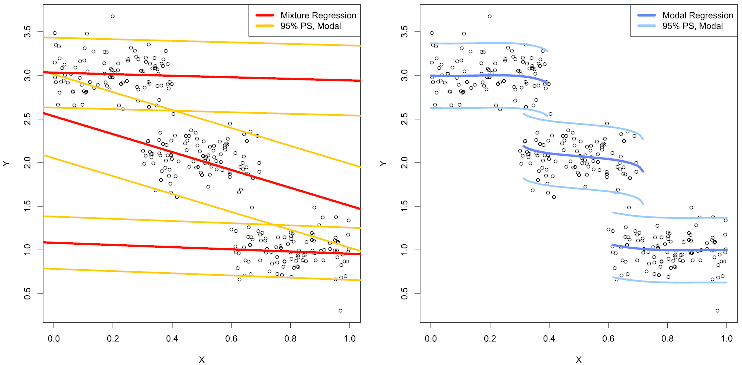}

\caption{A comparison between mixture regression, on the left, and
modal regression, on the right.}
\label{Figmix}
\end{figure}

Figure~\ref{Figmix} gives a comparison between linear mixture
regression and modal regression. We fit the linear mixture model using
the R package {\sf mixtools}, specifying $k=3$ components, over
10,000 runs of the EM algorithm (choosing eventually the result the
highest likelihood value). The modal regression estimate used a
bandwidth value that minimized the volume of the corresponding
prediction set, as characterized in Figure~\ref{Figh}. The figure
reveals yet another important difference between the two methods: the
estimated modal regression trends do not persist across the whole $x$
domain, while the linear mixture model (in its default specification)
carries the estimated linear trends across the entirety of the $x$
domain. This is due to assumption (MR2), which models each component
probability $\pi_j$ as a constant, independent of $x$. As a result,
the prediction set from the linear mixture model has a much larger
volume than that from modal regression, since it vacuously covers the
extensions of each linear fit across the whole domain. Relaxing
assumption (MR2) would address this issue, but it would also make
the mixture estimation more difficult.

%

\subsection{Clustering with modal regression}\label{secclustering}

We now describe how modal regression can be used to conduct
clustering, conditional on $x$. This clustering leads us
to define
modal proportions and modal dispersions, which are roughly
analogous to the component parameters $\pi_j(x)$ and
{$\sigma^2_j(x)$} in mixture regression.

Mode-based clustering
[\citet
{cheng1995mean,comaniciu2002mean,Li2007,yao2009bayesian,Chen2014EMC}]
is a nonparametric clustering method which uses
local density modes to define clusters.
A similar idea applies to modal regression.
In words, at each point $x$, we find the modes of $p(y\mid x)$ and
we cluster according to the basins of attractions of these modes.
Formally,
at each $(x,y)$, we define an ascending path by
\[
\gamma_{(x,y)}: \R^{+}\rightarrow\mathbb{K}\times\K, \qquad
\gamma_{(x,y)}(0) = (x,y),\qquad\gamma'_{(x,y)}(t) =
\bigl(0, p_y(x,y)\bigr).
\]
That is, $\gamma_{(x,y)}$ is the gradient ascent path
in the $y$ direction (with $x$ fixed), starting at the point $y$.
Denote the destination of the path by
$\dest(x,y) = \lim_{t\rightarrow\infty} \gamma_{(x,y)}(t)$.
By Morse theory, $\dest(x,y)=m_j(x)$ for one and only one regression
mode $m_j(x)$, $j=1,\ldots, K$. Thus, we assign the cluster label $j$
to the point $(x,y)$.
Similar ideas have been used by \citet{Li2007,yao2009bayesian}, and the
former authors also discuss how the modes and clustering results merge
as the bandwidth increases.

\begin{figure}[t]

\includegraphics{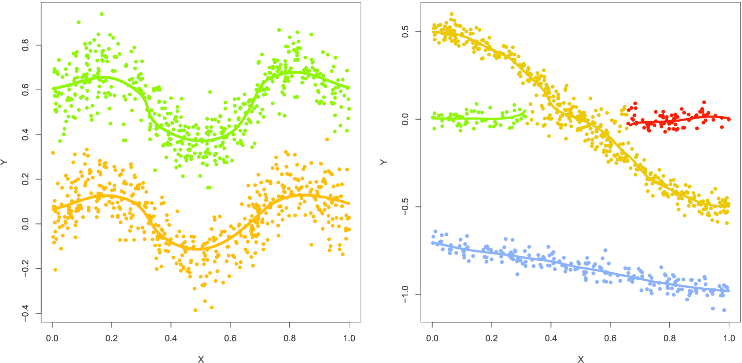}

\caption{Two examples of clustering based on modal regression.}
\label{Figcluster}
\end{figure}

The above was a population-level description of the clusters.
In practice, we use the mean-shift algorithm (Algorithm
\ref{algPMS}) to estimate clusters and assign points according
to the output of the algorithm. That is, by iterating
the mean-shift update~\eqref{eqms} for each\vspace*{1pt} point
$(X_i,Y_i)$, with $X_i$ fixed, we arrive at an estimated mode
{$\hat{m}_j(X_i)$} for some {$j=1,\ldots,\hat{K}$}, and
we hence assign $(X_i,Y_i)$ to cluster $j$. An issue is that
determination of the estimated modal functions
{$\hat{m}_j$, $j=1,\ldots, \hat{K}$}, or equivalently, of the
modal manifolds
{$\hat{S}_1,\ldots,\hat{S}_{\hat{K}}$}, is not immediate from
the data samples. These are well-defined in principle, but require
running the mean-shift algorithm at each input point $x$. In
data examples, therefore, we run mean-shift over a fine mesh (e.g.,\vspace*{1pt}
the data samples themselves) and apply hierarchical clustering to find
the collection {$\hat{S}_1,\ldots,\hat{S}_{\hat{K}}$}. It is
important to note that the latter clustering task, which seeks a
clustering of the outputs of the mean-shift algorithm, is trivial
compared to the original task (clustering of the data samples). Some
examples are shown in Figure~\ref{Figcluster}.

%

The clustering assignments give rise to the concepts of
modal proportions and modal dispersions. The modal proportion of
cluster $j$ is defined as
\[
\hat{q}_j = N_j/n,
\]
where {$N_j=\sum_{i=1}^n \mathbh{1}(i \in\hat{C}_j)$} is the\vspace*{2pt}
number of data points belonging to the $j$th cluster
{$\hat{C}_j$}. The modal dispersion of cluster $j$ is defined as
\[
\hat{\rho}^2_j = \frac{1}{N_j} \sum
_{i \in\hat{C}_j} \bigl(Y_i - \hat{m}(Y_i)
\bigr)^2,
\]
where $\hat{m}(Y_i)$ denotes the sample destination at $(X_i,Y_i)$
[i.e., the output of the mean-shift algorithm at $(X_i,Y_i)$]. This
is a measure of the spread of the data points around the $j$th
estimated modal manifold.

In a mixture regression model,
where each $\phi_j$ is assumed to be Gaussian,
the local modes of $p(y\mid x)$ behave like
the mixture centers $\mu_1(x),\ldots,\mu_K(x)$.
Thus, estimating the local modes is like
estimating the centers
of
the Gaussian mixtures.
The clustering based on modal regression is like
the recovery process for the mixing mechanism.
Each cluster can be thought of a mixture component
and hence the quantities {$\hat{q}_j, \hat{\rho}^2_j$}
are analogous to the estimates
{$\hat{\pi}_j, \hat{\sigma}_j^2$} in mixture regression
[assuming (MR2) and (MR4), so that to the mixture proportions and
variances do not depend on $x$].

\section{Comparison to density ridges}\label{secdensity}

Another concept related to modal regression estimation is that of
density ridge estimation. Relative to mixture regression, the
literature on density ridges is sparse; see
Chen, Genovese and Wasserman  (\citeyear{Chen2014GMRE,Chen2014ridge}),
\citet{Eberly1996},
\citet{genovese2014nonparametric}.

For simplicity of comparison, assume that the predictor $X$ is
univariate \mbox{($d=1$)}.
Let $v_1(x,y),v_2(x,y)$ be the eigenvectors
corresponding to the eigenvalues $\lambda_1(x,y)\geq\lambda_2(x,y)$
of $H(x,y)=\nabla^2 p(x,y)$, the Hessian matrix of density function
$p$ at $(x,y)$. Each point in the ridge set at $x$ is the local mode
of the local mode of subspace spanned by $v_2(x,y)$ with
$\lambda_{2}(x,y)<0$. We can express this as
\[
R(x) = \bigl\{ y: v_2(x,y)^T \nabla p(x,y) = 0,
v_2^T(x,y) H(x,y)v_2(x,y)<0 \bigr\}.
\]
Note that we can similarly express the modal set at $x$ as
\[
M(x) = \bigl\{y: 1_Y^T\nabla p(x,y)= 0,
1_Y^T H(x,y)1_Y<0\bigr\},
\]
where $1^T_Y= (0,1)$ is the unit vector in the $y$ direction.
As can be seen easily, the key difference lies in the two vectors
$1_Y$ and $v_2(x,y)$. Every point on the density ridge
is local mode with respect to a different subspace, while
every point on the modal regression is the local mode
with respect to the same subspace, namely, that aligned with the
$y$-axis. The following simple lemma describes cases in which these
two sets coincide.

%
\begin{lem}[(Equivalence of modal and ridge sets)]\label{lemMRDR}
Assume that $d=1$, fix any
point $x$, and let $y \in M(x)$. Then provided that:
\begin{longlist}[2.]
\item[1.] $p_x(x,y)=0$, or
\item[2.]$p_{xy}(x,y)=0$,
\end{longlist}
it also holds that $y \in R(x)$.
\end{lem}

The proof is in the supplementary material [\citet{Chen2015Modalsupp}].
The lemma asserts that a conditional mode where the
density is locally stationary, that is, $p_x(x, y) = 0$, or the density is
locally isotropic, that is, $p_{xy}(x,y)=0$, is also a density ridge.
More explicitly, the first condition states that saddle points and
local maximums are both local modes and ridge points,
and the second condition states that when modal manifolds moving
along the $x$-axis, they are also density ridges.

\begin{figure}

\includegraphics{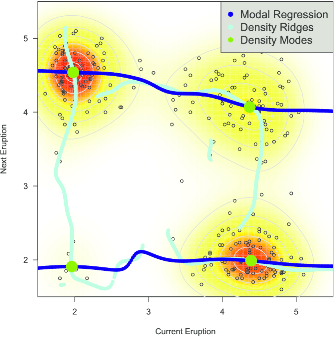}

\caption{A comparison between modal regression, density ridges
and density modes using the old faithful data set.
The background color represents the joint density (red:
high density).}
\label{Figcomp}
\end{figure}

We compare modal regression, density ridges, and density modes
in Figure~\ref{Figcomp}. Both the estimated density ridges and
modal manifolds pass through the density modes, as predicted by
Lemma~\ref{lemMRDR}. Furthermore, at places in which the joint
density is locally isotropic (i.e., spherical), the modal regression
and density ridge components roughly coincide.

%

From a general perspective, modal regression and density ridges
are looking for different types of structures; modal regression
examines the conditional structure of $Y\mid X$, and density ridges seek
out the joint structure of $X,Y$.
Typically, density ridge estimation is less stable than modal
regression estimation because in the former, both the modes and the
subspace of interest [the second eigenvector $v_2(x,y)$ of the local
Hessian] must be estimated.

\section{Discussion}\label{secdiscussion}

We have investigated a nonparametric method for modal regression
estimation, based on a KDE of a joint sample of data points
$(X_1,Y_1),\ldots,  (X_n,Y_n)$. We studied some of the geometry
underlying the modal regression set, and described techniques for
confidence set estimation, prediction set estimation, and
bandwidth selection for the underlying KDE. Finally, we compared the
proposed method to the well-studied mixture of regression model, and
the less well known but also highly relevant problem of density ridge
estimation. The main message is that nonparametric modal regression
offers a relatively simple and useable tool to capture conditional
structure missed by conventional regression methods. The
advances we have developed in this paper, such those for constructing
confidence sets and prediction sets, add to its usefulness as a
practical tool.

\begin{figure}

\includegraphics{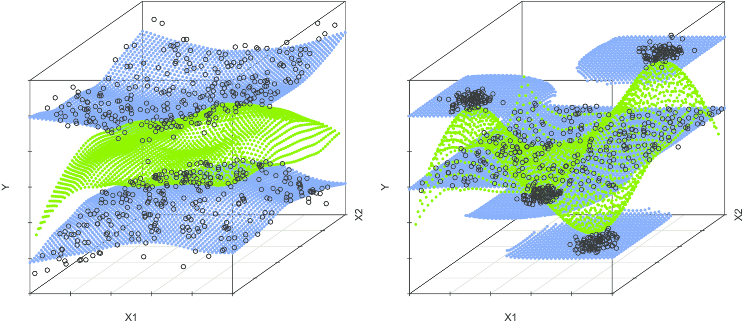}

\caption{Two examples for $d=2$. Modal regression estimates are
shown in blue, and local regression in green.}
\label{Fig3d}
\end{figure}

Though the discussion in this paper treated the dimension $d$ of the
predictor variable $X$ as arbitrary, all examples used $d=1$. We
finish by giving two simple examples for $d=2$.
In the first example, the data points are normally distributed around
two parabolic surfaces; in the second example, the data points come
from five different components of two-dimensional structure.
We apply both modal regression (in blue) and local regression (in
green) to the two examples, shown in Figure~\ref{Fig3d}. The
estimated modal regression set identifies the appropriate structure,
while local regression does not (most of the local regression surface
does not lie near any of the data points at all).

%



\section*{Acknowledgement}
We thank the reviewers for useful comments.

\begin{supplement}[id=suppA]
\sname{Supplementary Proofs}
\stitle{Nonparametric modal regression\\}
\slink[doi]{10.1214/15-AOS1373SUPP} 
\sdatatype{.pdf}
\sfilename{aos1373\_supp.pdf}
\sdescription{This document contains all proofs to the theorems and lemmas in this paper.}
\end{supplement}

%

\printaddresses

\begin{thebibliography}{36}

\bibitem[\protect\citeauthoryear{Arias-Castro, Mason and
Pelletier}{2013}]{arias2013estimation}
%
\begin{bmisc}[auto:parserefs-M02]
\bauthor{\bsnm{Arias-Castro},~\bfnm{E.}\binits{E.}},
\bauthor{\bsnm{Mason},~\bfnm{D.}\binits{D.}} \AND
\bauthor{\bsnm{Pelletier},~\bfnm{B.}\binits{B.}}
(\byear{2013}).
\btitle{On the estimation of the gradient lines of a density and the
consistency of the mean-shift algorithm}.
\bnote{Unpublished Manuscript}.
\end{bmisc}
%
\bptok{imsref}%
\endbibitem

\bibitem[\protect\citeauthoryear{Bishop}{2006}]{Bishop2006}
%
\begin{bbook}[mr]
\bauthor{\bsnm{Bishop},~\bfnm{Christopher~M.}\binits{C.~M.}}
(\byear{2006}).
\btitle{Pattern Recognition and Machine Learning}.
\bpublisher{Springer},
\blocation{New York}.
\bid{doi={10.1007/978-0-387-45528-0}, mr={2247587}}
\end{bbook}
%
\bptok{imsref}%
\endbibitem

\bibitem[\protect\citeauthoryear{Carreira-Perpi{\~{n}}{\'
{a}}n}{2007}]{carreira2007gaussian}
%
\begin{barticle}[auto:parserefs-M02]
\bauthor{\bsnm{Carreira-Perpi{\~{n}}{\'{a}}n},~\bfnm{M.~{\'
{A}}.}\binits{M.~\'{A}.}}
(\byear{2007}).
\btitle{Gaussian mean-shift is an em algorithm}.
\bjournal{IEEE Trans. Pattern Anal. Mach. Intell.}
\bvolume{29}
\bpages{0767--0776}.
\end{barticle}
%
\bptok{imsref}%
\endbibitem

\bibitem[\protect\citeauthoryear{Chaganty and Liang}{2013}]{chaganty2013spectral}
\begin{binproceedings}[auto:parserefs-M02]
\bauthor{\bsnm{Chaganty},~\bfnm{A.~T.}\binits{A.~T.}} \AND
\bauthor{\bsnm{Liang},~\bfnm{P.}\binits{P.}}
(\byear{2013}).
\btitle{Spectral experts for estimating mixtures of linear regressions}.
In \bbooktitle{Proceedings of the 30th International Conference on Machine Learning (ICML-13)}
\bpages{1040--1048}.
\bpublisher{ACM},
\blocation{New York}.
\end{binproceedings}
%
\bptok{imsref}%
\endbibitem

\bibitem[\protect\citeauthoryear{Chen, Genovese and Wasserman}{2014a}]{Chen2014EMC}
%
\begin{bmisc}[auto:parserefs-M02]
\bauthor{\bsnm{Chen},~\bfnm{Y.-C.}\binits{Y.-C.}},
\bauthor{\bsnm{Genovese},~\bfnm{C.~R.}\binits{C.~R.}} \AND
\bauthor{\bsnm{Wasserman},~\bfnm{L.}\binits{L.}}
(\byear{2014}a).
\bhowpublished{Enhanced mode clustering. Available at \arxivurl{arXiv:1406.1780}}.
\end{bmisc}
%
\bptok{imsref}%
\endbibitem

\bibitem[\protect\citeauthoryear{Chen, Genovese and
Wasserman}{2014b}]{Chen2014GMRE}
%
\begin{bmisc}[auto:parserefs-M02]
\bauthor{\bsnm{Chen},~\bfnm{Y.-C.}\binits{Y.-C.}},
\bauthor{\bsnm{Genovese},~\bfnm{C.~R.}\binits{C.~R.}} \AND
\bauthor{\bsnm{Wasserman},~\bfnm{L.}\binits{L.}}
(\byear{2014}b).
\bhowpublished{Generalized mode and ridge estimation.
Available at \arxivurl{arXiv:1406.1803}}.
\end{bmisc}
%
\bptok{imsref}%
\endbibitem

\bibitem[\protect\citeauthoryear{Chen, Genovese and Wasserman}{2015}]{Chen2014ridge}
%
\begin{barticle}[mr]
\bauthor{\bsnm{Chen},~\bfnm{Yen-Chi}\binits{Y.-C.}},
\bauthor{\bsnm{Genovese},~\bfnm{Christopher~R.}\binits{C.~R.}} \AND
\bauthor{\bsnm{Wasserman},~\bfnm{Larry}\binits{L.}}
(\byear{2015}).
\btitle{Asymptotic theory for density ridges}.
\bjournal{Ann. Statist.}
\bvolume{43}
\bpages{1896--1928}.
\bid{doi={10.1214/15-AOS1329}, issn={0090-5364}, mr={3375871}}
\bptnote{check volume, check pages, check year}%
\end{barticle}
%
\bptok{imsref}%
\endbibitem

\bibitem[\protect\citeauthoryear{Chen et~al.}{2015}]{Chen2015Modalsupp}
%
\begin{bmisc}[auto]
\bauthor{\bsnm{Chen},~\bfnm{Y.-C.}\binits{Y.-C.}},
\bauthor{\bsnm{Genovese},~\bfnm{C.~R.}\binits{C.~R.}},
\bauthor{\bsnm{Tibshirani},~\bfnm{R.~J.}\binits{R.~J.}} \AND
\bauthor{\bsnm{Wasserman},~\bfnm{L.}\binits{L.}}
(\byear{2015}).
\bhowpublished{Supplement to ``Nonparametric modal regression.''
DOI:\doiurl{10.1214/15-AOS1373SUPP}}.
\end{bmisc}
%
\bptok{imsref}%
\endbibitem

\bibitem[\protect\citeauthoryear{Cheng}{1995}]{cheng1995mean}
%
\begin{barticle}[auto:parserefs-M02]
\bauthor{\bsnm{Cheng},~\bfnm{Y.}\binits{Y.}}
(\byear{1995}).
\btitle{Mean shift, mode seeking, and clustering}.
\bjournal{IEEE Trans. Pattern Anal. Mach. Intell.}
\bvolume{17}
\bpages{790--799}.
\end{barticle}
%
\bptok{imsref}%
\endbibitem

\bibitem[\protect\citeauthoryear{Chernozhukov, Chetverikov and Kato}{2014a}]{chernozhukov2014anti}
%
\begin{barticle}[mr]
\bauthor{\bsnm{Chernozhukov},~\bfnm{Victor}\binits{V.}},
\bauthor{\bsnm{Chetverikov},~\bfnm{Denis}\binits{D.}} \AND
\bauthor{\bsnm{Kato},~\bfnm{Kengo}\binits{K.}}
(\byear{2014}a).
\btitle{Anti-concentration and honest, adaptive confidence bands}.
\bjournal{Ann. Statist.}
\bvolume{42}
\bpages{1787--1818}.
\bid{doi={10.1214/14-AOS1235}, issn={0090-5364}, mr={3262468}}
\end{barticle}
%
\bptok{imsref}%
\endbibitem

\bibitem[\protect\citeauthoryear{Chernozhukov, Chetverikov and
Kato}{2014b}]{chernozhukov2014gaussian}
%
\begin{barticle}[mr]
\bauthor{\bsnm{Chernozhukov},~\bfnm{Victor}\binits{V.}},
\bauthor{\bsnm{Chetverikov},~\bfnm{Denis}\binits{D.}} \AND
\bauthor{\bsnm{Kato},~\bfnm{Kengo}\binits{K.}}
(\byear{2014}b).
\btitle{Gaussian approximation of suprema of empirical processes}.
\bjournal{Ann. Statist.}
\bvolume{42}
\bpages{1564--1597}.
\bid{doi={10.1214/14-AOS1230}, issn={0090-5364}, mr={3262461}}
\end{barticle}
%
\bptok{imsref}%
\endbibitem

\bibitem[\protect\citeauthoryear{Comaniciu and
Meer}{2002}]{comaniciu2002mean}
%
\begin{barticle}[auto:parserefs-M02]
\bauthor{\bsnm{Comaniciu},~\bfnm{D.}\binits{D.}} \AND
\bauthor{\bsnm{Meer},~\bfnm{P.}\binits{P.}}
(\byear{2002}).
\btitle{Mean shift: A robust approach toward feature space analysis}.
\bjournal{IEEE Trans. Pattern Anal. Mach. Intell.}
\bvolume{24}
\bpages{603--619}.
\end{barticle}
%
\bptok{imsref}%
\endbibitem

\bibitem[\protect\citeauthoryear{Eberly}{1996}]{Eberly1996}
%
\begin{bbook}[auto:parserefs-M02]
\bauthor{\bsnm{Eberly},~\bfnm{D.}\binits{D.}}
(\byear{1996}).
\btitle{Ridges in Image and Data Analysis}.
\bpublisher{Springer},
\blocation{Berlin}.
\end{bbook}
%
\bptok{imsref}%
\endbibitem

\bibitem[\protect\citeauthoryear{Efron}{1979}]{Efron1979}
%
\begin{barticle}[mr]
\bauthor{\bsnm{Efron},~\bfnm{B.}\binits{B.}}
(\byear{1979}).
\btitle{Bootstrap methods: Another look at the jackknife}.
\bjournal{Ann. Statist.}
\bvolume{7}
\bpages{1--26}.
\bid{issn={0090-5364}, mr={0515681}}
\end{barticle}
%
\bptok{imsref}%
\endbibitem

\bibitem[\protect\citeauthoryear{Einbeck and Tutz}{2006}]{Einbeck2006}
%
\begin{barticle}[mr]
\bauthor{\bsnm{Einbeck},~\bfnm{Jochen}\binits{J.}} \AND
\bauthor{\bsnm{Tutz},~\bfnm{Gerhard}\binits{G.}}
(\byear{2006}).
\btitle{Modelling beyond regression functions: An application of
multimodal regression to speed-flow data}.
\bjournal{J. Roy. Statist. Soc. Ser. C}
\bvolume{55}
\bpages{461--475}.
\bid{doi={10.1111/j.1467-9876.2006.00547.x}, issn={0035-9254}, mr={2242274}}
\end{barticle}
%
\bptok{imsref}%
\endbibitem

\bibitem[\protect\citeauthoryear{Einmahl and
Mason}{2005}]{einmahl2005uniform}
%
\begin{barticle}[mr]
\bauthor{\bsnm{Einmahl},~\bfnm{Uwe}\binits{U.}} \AND
\bauthor{\bsnm{Mason},~\bfnm{David~M.}\binits{D.~M.}}
(\byear{2005}).
\btitle{Uniform in bandwidth consistency of kernel-type function estimators}.
\bjournal{Ann. Statist.}
\bvolume{33}
\bpages{1380--1403}.
\bid{doi={10.1214/009053605000000129}, issn={0090-5364}, mr={2195639}}
\end{barticle}
%
\bptok{imsref}%
\endbibitem

\bibitem[\protect\citeauthoryear{Genovese
et~al.}{2014}]{genovese2014nonparametric}
%
\begin{barticle}[mr]
\bauthor{\bsnm{Genovese},~\bfnm{Christopher~R.}\binits{C.~R.}},
\bauthor{\bsnm{Perone-Pacifico},~\bfnm{Marco}\binits{M.}},
\bauthor{\bsnm{Verdinelli},~\bfnm{Isabella}\binits{I.}} \AND
\bauthor{\bsnm{Wasserman},~\bfnm{Larry}\binits{L.}}
(\byear{2014}).
\btitle{Nonparametric ridge estimation}.
\bjournal{Ann. Statist.}
\bvolume{42}
\bpages{1511--1545}.
\bid{doi={10.1214/14-AOS1218}, issn={0090-5364}, mr={3262459}}
\end{barticle}
%
\bptok{imsref}%
\endbibitem

\bibitem[\protect\citeauthoryear{Gin{\'e} and Guillou}{2002}]{gine2002rates}
%
\begin{barticle}[mr]
\bauthor{\bsnm{Gin{\'e}},~\bfnm{Evarist}\binits{E.}} \AND
\bauthor{\bsnm{Guillou},~\bfnm{Armelle}\binits{A.}}
(\byear{2002}).
\btitle{Rates of strong uniform consistency for multivariate kernel
density estimators}.
\bjournal{Ann. Inst. Henri Poincar\'e Probab. Stat.}
\bvolume{38}
\bpages{907--921}.
\bid{doi={10.1016/S0246-0203(02)01128-7}, issn={0246-0203}, mr={1955344}}
\end{barticle}
%
\bptok{imsref}%
\endbibitem

\bibitem[\protect\citeauthoryear{Huang, Li and Wang}{2013}]{Huang2013jasa}
%
\begin{barticle}[mr]
\bauthor{\bsnm{Huang},~\bfnm{Mian}\binits{M.}},
\bauthor{\bsnm{Li},~\bfnm{Runze}\binits{R.}} \AND
\bauthor{\bsnm{Wang},~\bfnm{Shaoli}\binits{S.}}
(\byear{2013}).
\btitle{Nonparametric mixture of regression models}.
\bjournal{J. Amer. Statist. Assoc.}
\bvolume{108}
\bpages{929--941}.
\bid{doi={10.1080/01621459.2013.772897}, issn={0162-1459}, mr={3174674}}
\end{barticle}
%
\bptok{imsref}%
\endbibitem

\bibitem[\protect\citeauthoryear{Huang and Yao}{2012}]{Huang2012jasa}
%
\begin{barticle}[mr]
\bauthor{\bsnm{Huang},~\bfnm{Mian}\binits{M.}} \AND
\bauthor{\bsnm{Yao},~\bfnm{Weixin}\binits{W.}}
(\byear{2012}).
\btitle{Mixture of regression models with varying mixing proportions:
A semiparametric approach}.
\bjournal{J. Amer. Statist. Assoc.}
\bvolume{107}
\bpages{711--724}.
\bid{doi={10.1080/01621459.2012.682541}, issn={0162-1459}, mr={2980079}}
\end{barticle}
%
\bptok{imsref}%
\endbibitem

\bibitem[\protect\citeauthoryear{Hunter and
Young}{2012}]{hunter2012semiparametric}
%
\begin{barticle}[mr]
\bauthor{\bsnm{Hunter},~\bfnm{David~R.}\binits{D.~R.}} \AND
\bauthor{\bsnm{Young},~\bfnm{Derek~S.}\binits{D.~S.}}
(\byear{2012}).
\btitle{Semiparametric mixtures of regressions}.
\bjournal{J.~Nonparametr. Stat.}
\bvolume{24}
\bpages{19--38}.
\bid{doi={10.1080/10485252.2011.608430}, issn={1048-5252}, mr={2885823}}
\end{barticle}
%
\bptok{imsref}%
\endbibitem

\bibitem[\protect\citeauthoryear{Hyndman, Bashtannyk and
Grunwald}{1996}]{Hyndman1996}
%
\begin{barticle}[mr]
\bauthor{\bsnm{Hyndman},~\bfnm{Rob~J.}\binits{R.~J.}},
\bauthor{\bsnm{Bashtannyk},~\bfnm{David~M.}\binits{D.~M.}} \AND
\bauthor{\bsnm{Grunwald},~\bfnm{Gary~K.}\binits{G.~K.}}
(\byear{1996}).
\btitle{Estimating and visualizing conditional densities}.
\bjournal{J. Comput. Graph. Statist.}
\bvolume{5}
\bpages{315--336}.
\bid{doi={10.2307/1390887}, issn={1061-8600}, mr={1422114}}
\end{barticle}
%
\bptok{imsref}%
\endbibitem

\bibitem[\protect\citeauthoryear{Jacobs et~al.}{1991}]{Jacobs1991}
%
\begin{barticle}[auto:parserefs-M02]
\bauthor{\bsnm{Jacobs},~\bfnm{R.~A.}\binits{R.~A.}},
\bauthor{\bsnm{Jordan},~\bfnm{M.~I.}\binits{M.~I.}},
\bauthor{\bsnm{Nowlan},~\bfnm{S.~J.}\binits{S.~J.}} \AND
\bauthor{\bsnm{Hinton},~\bfnm{G.~E.}\binits{G.~E.}}
(\byear{1991}).
\btitle{Adaptive mixtures of local experts}.
\bjournal{Neural Comput.}
\bvolume{3}
\bpages{79--87}.
\bnote{ISSN 0899-7667. Available at
\surl{http://dx.doi.org/10.1162/neco.1991.3.1.79}}.
\end{barticle}
%
\bptok{imsref}%
\endbibitem

\bibitem[\protect\citeauthoryear{Jiang and Tanner}{1999}]{Jiang1999}
%
\begin{barticle}[mr]
\bauthor{\bsnm{Jiang},~\bfnm{Wenxin}\binits{W.}} \AND
\bauthor{\bsnm{Tanner},~\bfnm{Martin~A.}\binits{M.~A.}}
(\byear{1999}).
\btitle{Hierarchical mixtures-of-experts for exponential family
regression models: Approximation and maximum likelihood estimation}.
\bjournal{Ann. Statist.}
\bvolume{27}
\bpages{987--1011}.
\bid{doi={10.1214/aos/1018031265}, issn={0090-5364}, mr={1724038}}
\end{barticle}
%
\bptok{imsref}%
\endbibitem

\bibitem[\protect\citeauthoryear{Khalili and Chen}{2007}]{Khalili2007jasa}
%
\begin{barticle}[mr]
\bauthor{\bsnm{Khalili},~\bfnm{Abbas}\binits{A.}} \AND
\bauthor{\bsnm{Chen},~\bfnm{Jiahua}\binits{J.}}
(\byear{2007}).
\btitle{Variable selection in finite mixture of regression models}.
\bjournal{J. Amer. Statist. Assoc.}
\bvolume{102}
\bpages{1025--1038}.
\bid{doi={10.1198/016214507000000590}, issn={0162-1459}, mr={2411662}}
\end{barticle}
%
\bptok{imsref}%
\endbibitem

\bibitem[\protect\citeauthoryear{Lee}{1989}]{Lee1989}
%
\begin{barticle}[mr]
\bauthor{\bsnm{Lee},~\bfnm{Myoung-jae}\binits{M.-j.}}
(\byear{1989}).
\btitle{Mode regression}.
\bjournal{J. Econometrics}
\bvolume{42}
\bpages{337--349}.
\bid{doi={10.1016/0304-4076(89)90057-2}, issn={0304-4076}, mr={1040748}}
\end{barticle}
%
\bptok{imsref}%
\endbibitem

\bibitem[\protect\citeauthoryear{Li, Ray and Lindsay}{2007}]{Li2007}
%
\begin{barticle}[mr]
\bauthor{\bsnm{Li},~\bfnm{Jia}\binits{J.}},
\bauthor{\bsnm{Ray},~\bfnm{Surajit}\binits{S.}} \AND
\bauthor{\bsnm{Lindsay},~\bfnm{Bruce~G.}\binits{B.~G.}}
(\byear{2007}).
\btitle{A nonparametric statistical approach to clustering via mode
identification}.
\bjournal{J. Mach. Learn. Res.}
\bvolume{8}
\bpages{1687--1723}.
\bid{issn={1532-4435}, mr={2332445}}
\end{barticle}
%
\bptok{imsref}%
\endbibitem

\bibitem[\protect\citeauthoryear{Rojas}{2005}]{Rojas2005}
%
\begin{bmisc}[auto:parserefs-M02]
\bauthor{\bsnm{Rojas},~\bfnm{A.}\binits{A.}}
(\byear{2005}).
\bhowpublished{Nonparametric mixture regression.
Ph.D. thesis, Carnegie Mellon Univ., Pittsburgh, PA.}
\end{bmisc}
%
\bptok{imsref}%
\endbibitem

\bibitem[\protect\citeauthoryear{Romano}{1988}]{romano1988weak}
%
\begin{barticle}[mr]
\bauthor{\bsnm{Romano},~\bfnm{Joseph~P.}\binits{J.~P.}}
(\byear{1988}).
\btitle{On weak convergence and optimality of kernel density estimates
of the mode}.
\bjournal{Ann. Statist.}
\bvolume{16}
\bpages{629--647}.
\bid{doi={10.1214/aos/1176350824}, issn={0090-5364}, mr={0947566}}
\bptnote{check volume}%
\end{barticle}
%
\bptok{imsref}%
\endbibitem

\bibitem[\protect\citeauthoryear{Sager and Thisted}{1982}]{sager1982maximum}
%
\begin{barticle}[mr]
\bauthor{\bsnm{Sager},~\bfnm{Thomas~W.}\binits{T.~W.}} \AND
\bauthor{\bsnm{Thisted},~\bfnm{Ronald~A.}\binits{R.~A.}}
(\byear{1982}).
\btitle{Maximum likelihood estimation of isotonic modal regression}.
\bjournal{Ann. Statist.}
\bvolume{10}
\bpages{690--707}.
\bid{issn={0090-5364}, mr={0663426}}
\bptnote{check volume}%
\end{barticle}
%
\bptok{imsref}%
\endbibitem

\bibitem[\protect\citeauthoryear{Scott}{1992}]{scott1992multivariate}
%
\begin{bbook}[mr]
\bauthor{\bsnm{Scott},~\bfnm{David~W.}\binits{D.~W.}}
(\byear{1992}).
\btitle{Multivariate Density Estimation: Theory, Practice, and Visualization}.
\bpublisher{Wiley},
\blocation{New York}.
\bid{doi={10.1002/9780470316849}, mr={1191168}}
\end{bbook}
%
\bptok{imsref}%
\endbibitem

\bibitem[\protect\citeauthoryear{Viele and Tong}{2002}]{viele2002modeling}
%
\begin{barticle}[mr]
\bauthor{\bsnm{Viele},~\bfnm{Kert}\binits{K.}} \AND
\bauthor{\bsnm{Tong},~\bfnm{Barbara}\binits{B.}}
(\byear{2002}).
\btitle{Modeling with mixtures of linear regressions}.
\bjournal{Stat. Comput.}
\bvolume{12}
\bpages{315--330}.
\bid{doi={10.1023/A:1020779827503}, issn={0960-3174}, mr={1951705}}
\end{barticle}
%
\bptok{imsref}%
\endbibitem

\bibitem[\protect\citeauthoryear{Yao}{2013}]{yao2013note}
%
\begin{barticle}[mr]
\bauthor{\bsnm{Yao},~\bfnm{Weixin}\binits{W.}}
(\byear{2013}).
\btitle{A note on EM algorithm for mixture models}.
\bjournal{Statist. Probab. Lett.}
\bvolume{83}
\bpages{519--526}.
\bid{doi={10.1016/j.spl.2012.10.017}, issn={0167-7152}, mr={3006984}}
\end{barticle}
%
\bptok{imsref}%
\endbibitem

\bibitem[\protect\citeauthoryear{Yao and Li}{2014}]{yao2014new}
%
\begin{barticle}[mr]
\bauthor{\bsnm{Yao},~\bfnm{Weixin}\binits{W.}} \AND
\bauthor{\bsnm{Li},~\bfnm{Longhai}\binits{L.}}
(\byear{2014}).
\btitle{A new regression model: Modal linear regression}.
\bjournal{Scand. J. Stat.}
\bvolume{41}
\bpages{656--671}.
\bid{doi={10.1111/sjos.12054}, issn={0303-6898}, mr={3249422}}
\end{barticle}
%
\bptok{imsref}%
\endbibitem

\bibitem[\protect\citeauthoryear{Yao and Lindsay}{2009}]{yao2009bayesian}
%
\begin{barticle}[mr]
\bauthor{\bsnm{Yao},~\bfnm{Weixin}\binits{W.}} \AND
\bauthor{\bsnm{Lindsay},~\bfnm{Bruce~G.}\binits{B.~G.}}
(\byear{2009}).
\btitle{Bayesian mixture labeling by highest posterior density}.
\bjournal{J.~Amer. Statist. Assoc.}
\bvolume{104}
\bpages{758--767}.
\bid{doi={10.1198/jasa.2009.0237}, issn={0162-1459}, mr={2751453}}
\bptnote{check pages}%
\end{barticle}
%
\bptok{imsref}%
\endbibitem

\bibitem[\protect\citeauthoryear{Yao, Lindsay and Li}{2012}]{Yao2012}
%
\begin{barticle}[mr]
\bauthor{\bsnm{Yao},~\bfnm{Weixin}\binits{W.}},
\bauthor{\bsnm{Lindsay},~\bfnm{Bruce~G.}\binits{B.~G.}} \AND
\bauthor{\bsnm{Li},~\bfnm{Runze}\binits{R.}}
(\byear{2012}).
\btitle{Local modal regression}.
\bjournal{J. Nonparametr. Stat.}
\bvolume{24}
\bpages{647--663}.
\bid{doi={10.1080/10485252.2012.678848}, issn={1048-5252}, mr={2968894}}
\end{barticle}
%
\bptok{imsref}%
\endbibitem

\end{thebibliography}
\end{document}